\documentclass[aps,prb,twocolumn,amsmath,amssymb,amsfonts,longbibliography,
    superscriptaddress,nobibnotes]{revtex4-2} 
\usepackage[utf8]{inputenc}
\usepackage[T1]{fontenc} 
\usepackage{ebgaramond}
\usepackage{graphicx} 
\usepackage{bm} 
\usepackage{times}
\usepackage{mathptmx}
\usepackage[
hypertexnames=false,
colorlinks=true,
citecolor=blue,
linkcolor=blue,
urlcolor=blue]
{hyperref}
\usepackage{soul}
\usepackage{float}

\usepackage{xcolor}

\graphicspath{{fig/}{./fig/}{.}}

\begin{document}

\title{Identification of the Majorana edge modes in tight-binding systems based on the Krylov method}

\author{Andrzej Więckowski} % AW: w 2020 roku korzytamy z utf8 ęąłóżź %AP hahahaaaaaaaaA nie % AP i dlatego w APS wyswietlales sie jak wickowski %nieprawda, to jest kwestia wpisania nazwiska w formularz w APS z wykorzystaniem boomerowego formatowania \k{e} its, to jest oddzielny mechanizmi od texa papieru %sie ciesz ze nie \c{e}
% \c{e} to jest największa patola, to jest hiszpanska i innych takich literka, pamiętam jak czyzu na wykladach tego uzywal juz wtedy mnie to koliło w oczy
% beka zostawmy te komenty na arXiV XD
\email[e-mail: ]{andrzej.wieckowski@pwr.edu.pl}
\affiliation{Department of Theoretical Physics, 
Faculty of Fundamental Problems of Technology,
Wrocław University of Science and Technology,
PL-50370 Wrocław, Poland}

\author{Andrzej Ptok}  
\email[e-mail: ]{aptok@mmj.pl}
\affiliation{Institute of Nuclear Physics, Polish Academy of Sciences, 
ul. W. E. Radzikowskiego 152, 31-342 Kraków, Poland}

\author{Marcin Mierzejewski}
\email[e-mail: ]{marcin.mierzejewski@pwr.edu.pl}
\affiliation{Department of Theoretical Physics, 
Faculty of Fundamental Problems of Technology,
Wrocław University of Science and Technology,
PL-50370 Wrocław, Poland}

\author{Michał Kupczyński}
\email[e-mail: ]{michal.kupczynski@pwr.edu.pl}
\affiliation{Department of Theoretical Physics, 
Faculty of Fundamental Problems of Technology,
Wrocław University of Science and Technology,
PL-50370 Wrocław, Poland}

\date{\today}

\begin{abstract}
Low dimensional structures in the non-trivial topological phase can host the in-gap Majorana bound states, identified experimentally as zero-bias peaks in differential conductance.
Theoretical methods for studying Majorana modes are mostly based on the bulk-boundary correspondence or exact diagonalization of finite systems via, e.g., Bogoliubov--de~Gennes formalism. 
In this paper, we develop an efficient method for identifying the Majorana in-gap (edge) states via looking for extreme eigenvalues of symmetric matrices.
The presented approach is based on the Krylov method and allows for study the spatial profile of the modes as well as the spectrum of the system.
The advantage of this method is the calculation cost, which shows linear dependence on the number of lattice sites.
The latter problem may be solved for very large clusters of arbitrary shape/geometry. 
In order to demonstrate the efficiency of our approach, we study two- and three-dimensional clusters described by the Kitaev and Rashba models for which we determine the number of Majorana modes and calculate their spatial structures.
Additionally, we discuss the impact of the system size on the physical properties of the topological phase of the magnetic nanoisland deposited on the superconducting surface.
In this case, we have shown that the eigenvalues of the in-gap states depend on the length of the system edge.
\end{abstract}

\maketitle

%%%%%%%%%%%%%%%%%%%%%%%%%%%%%%%%%%%%%%%%%%%%%
%%%%%%%%%%%%%%%%%%%%%%%%%%%%%%%%%%%%%%%%%%%%%
%%%%%%%%%%%%%%%%%%%%%%%%%%%%%%%%%%%%%%%%%%%%%

\section{Introduction}
\label{sec.intro}

In 2001 Kitaev proposed realization of the Majorana bound states
at the ends of a one-dimensional (1D) chain of spinless fermions with an inter-site paring~\cite{kitaev.01}.
This preeminent idea opened a period of the experimental and theoretical studies of these topological bound states~\cite{aguado.17,pawlak.hoffman.19,prada.sanjose.19,lutchyn.bakkers.18}.
One of the main properties of the Majorana zero modes (MZMs) is their non-Abelian statistics~\cite{nayak.simon.08}.
This makes MZMs a very attractive subject of study in the context of topological quantum computing~\cite{akhmerov.10,liu.wong.14,dassarma.freedman.15,aasen.hell.16,hoffman.schrade.16,alicea.oreg.11} and its practical implementation via {\it braiding protocols}~\cite{alicea.oreg.11,kim.tewari.15,beenakker.19,wieckowski.mierzejewski.20,trif.simon.19}.

\begin{figure}[!b]
    \centering
    \includegraphics[width=\columnwidth]{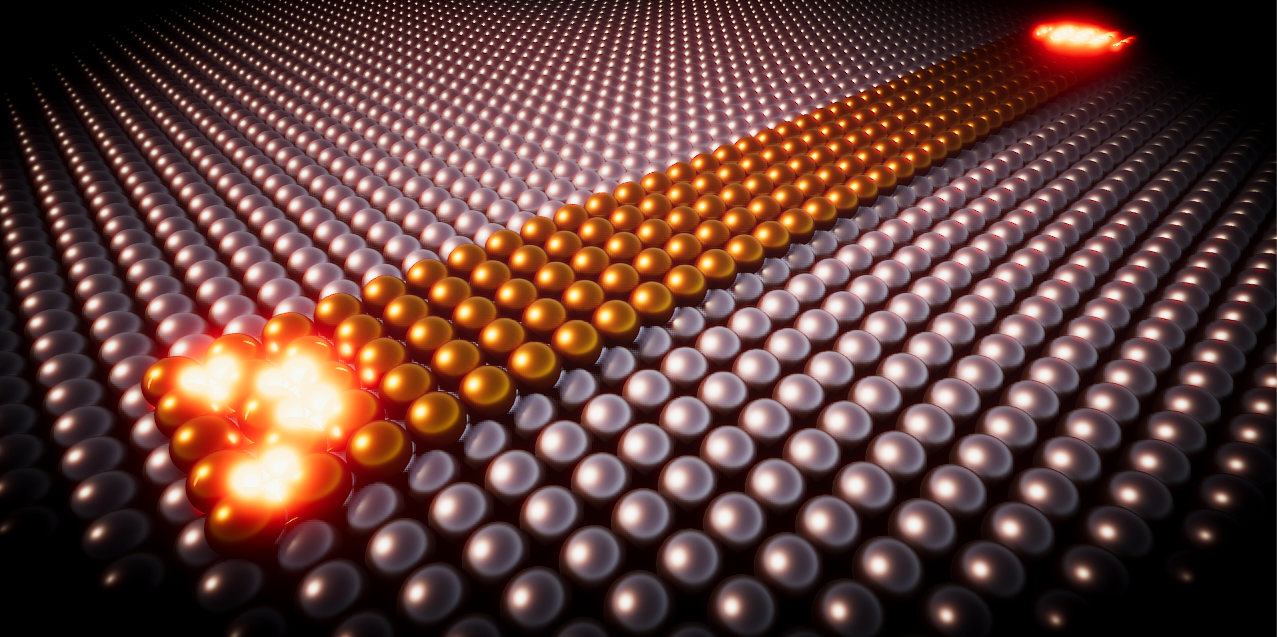}
    \caption{
    Schematic representation realization of the Majorana modes on the edge of the quasi-one-dimensional system deposited on the superconducting substrate.
    \label{fig.schemat}
    }
\end{figure}

Currently, there are known several experimental setups where the Majorana quasiparticles realization is expected in low dimensional systems (Fig.~\ref{fig.schemat}).
We can mention here semiconducting--superconducting hybrid nanostructures~\cite{deng.yu.12,mourik.zuo.12,das.ronen.12,finck.vanharlingen.13,
nichele.drachmann.17,gul.zhang.18,deng.vaitiekenas.16,deng.vaitiekenas.18} or chains of magnetic atoms deposited on a superconducting surface~\cite{nadjperge.drozdov.14,pawlak.kisiel.16,feldman.randeria.16,ruby.heinrich.17,jeon.xie.17,kim.palaciomorales.18}.
In the first case, the mutual interplay between intrinsic spin--orbit coupling, superconducting proximity effect, and external Zeeman magnetic field leads to the realization of zero-energy bound states~\cite{lutchyn.bakkers.18}.
In the second case, Majorana quasiparticles are formed as a consequence of the magnetic moments order in ferromagnetic atomic chains~\cite{braunecker.simon.13,klinovaja.stano.13,vazifeh.franz.13,braunecker.simon.15,kaladzhyan.simon.17,andolina.simon.17}.

Experimental works also report the existence of the Majorana in-gap (edge) modes around two-dimensional topological superconducting domains~\cite{drozdov.alexandradinata.14,menard.guissart.17,palaciomorales.mascot.19} or magnetic nanoisland~\cite{rontynen.ojanen.15,li.neupert.16,rachel.mascot.17,jack.xie.19}.
Moreover, the topological properties of such systems can be tuned by the attached nanowire to the two dimensional (2D) plaquette~\cite{kobialka.domanski.19,mascot.cocklin.19}.
Also recent progress in the preparation of the nanostructures opens way to realization of the topological edge modes in a two-dimensional system ~\cite{drost.ojanen.17,girvosky.lado.17,yan.liljeroth.19}.
This possibility has been recently reported in the case of ferromagnet--superconductor heterostructures~\cite{kezilebieke.nurulhuda.19,kezilebieke.nurulhuda.20}.
In such cases, the Majorana flat band can be realized~\cite{sedlmayer.aguiarhualde.15,glodzik.ojanen.20} at the edge of the finite-size system~\cite{kaladzhyan.bena.17,duncan.manna.20}.

\paragraph*{Motivation.}
Several experimental methods are used to confirm the presence of the Majorana edge modes.
Most of them are based on the quantization of the differential conductance~\cite{zhang.liu.18}, showing the Majorana quasiparticles signatures in the form of the zero-bias peak~\cite{zhang.liu.19}.
There also exist a few theoretical methods to confirm the existence of the topological phase in the system~\cite{hasan.kane.10,bansil.lin.16,wen.17}.
The most direct approach is to calculate the topological number which type depends on the system symmetry class~\cite{altland.zirnbauer.97,schnyder.ryu.08,ryu.andreas.10}.
The study of the topological phase diagram based on the bulk--boundary correspondence~\cite{mong.shivamoggi.11,fukui.shiozaki.12}, is also available.
In practice, these techniques are applicable to systems which fulfill several conditions concerning e.g. the periodicity (relatively small unit cell of the system because reciprocal space description is needed), homogeneity (absence of disorder), etc.
Therefore, to take these into account the real space calculations should be performed.
In the paper, we derive an efficient general technique to study the Majorana edge modes in a wide class of non-interacting systems described by tight-binding models, independently of the translational symmetry, geometry or dimensionality.

The presented method can be used to study the MZMs realization in the system.
The method is based on the fact that the MZMs operators are integrals of motion of the system~\cite{lee.wilczek.13,dassarma.freedman.15,wieckowski.maska.18,monthus.18,wieckowski.ptok.19,wieckowski.mierzejewski.20}.
The number of the MZMs as well as their spatial structure may be obtained from the analysis of the extreme eigenvalues of an appropriate symmetric matrix.
As a consequence, algorithms based on the Krylov subspace-based decomposition can be used~\cite{saad.2003}.
By using those algorithms, our method allows for studying much larger systems in comparison to methods requiring full diagonalization of the Hamiltonian.
Moreover, this method can be generalized to study the Majorana states. 
We show that the method correctly reproduced the spectrum of the system, also in the absence of zero-energy MZMs.
What is important, the computational cost of numerical calculations depends linearly on the size of the system.
This advantage of our method opens new possibilities for the numerical studies of the in-gap (edge) states, including investigation of more realistic multi-orbital models or systems with complex spatial structures. 
One may also go beyond the simplest modeling of nanowires which host MZMs and describe them as real three-dimensional (3D) objects, as schematically shown in Fig.~\ref{fig.schemat} and discussed later on in more details.

The paper is organized as follows.
First, in Sec.~\ref{sec.method} we derive the algorithm.
We show analytically, that the presented method correctly reproduces the spectrum of the system.
Additionally, we discuss the numerical implementation and performance evaluation of the discussed method.
The basic examples and tests for the Rashba chain are shown in Sec.~\ref{sec.basic_example}.
In Sec.~\ref{sec.adv_example} we compare results obtained within the presented method applied to the systems and models well established in the literature.
In particular, we reproduce the physical behaviors of the MZMs in the case of the two-dimensional Kitaev model (Sec.~\ref{sec.2d_kitaev}), the two-dimensional Rashba stripe (Sec.~\ref{sec.2d_rashba}), and the three-dimensional Rashba wire (Sec.~\ref{sec.3d_rashba}).
Next, in Sec.~\ref{sec.island} we present original results for the magnetic nanoisland, i.e., the system with irregular shape and site-dependent quantities.
We discuss here, the scaling of the in-gap states with the size of the system.
Finally, we summarize our method and results in Sec.~\ref{sec.summary}.

%%%%%%%%%%%%%%%%%%%%%%%%%%%%%%%%%%%%%%%%%%%%%
%%%%%%%%%%%%%%%%%%%%%%%%%%%%%%%%%%%%%%%%%%%%%
%%%%%%%%%%%%%%%%%%%%%%%%%%%%%%%%%%%%%%%%%%%%%

\section{Method}
\label{sec.method}

\paragraph*{Main aspects.}
%Before sketching the new method, 
First we define Majorana fermions and MZMs. 
Majorana operators are defined by the following commutation relation
\begin{eqnarray}
    \left\{ \gamma_i,\gamma_j \right\} = 2\delta_{ij},
\end{eqnarray}
which lead to the following properties of Majorana fermions
\begin{eqnarray}
    \gamma_i^2 = 1,\quad \gamma_i=\gamma_i^\dagger.
\end{eqnarray}
 Majorana zero modes, $\Gamma_n$, are states  which are indistinguishable from their antistates, so they fulfill mentioned above conditions for Majorana fermions, and additionally, they have to commute with the Hamiltonian $H$~\cite{dassarma.freedman.15,wieckowski.maska.18},
\begin{equation}
[H,\Gamma_n] = 0.
\end{equation}

Every fermionic Hamiltonian, without many-body interaction, can be written in terms of Majorana operators
\begin{eqnarray}
\label{eq:general_Majorana_hamiltonian}
    H = \mathrm i \sum_{ij} M_{ij} \gamma_i\, \gamma_j.
\end{eqnarray}
The matrix $M$ has to be real and it can be always chosen in the upper triangular form, which is proven by the commutation relations for Majorana operators (see Appendix~\ref{app.details.M_matrix}).

Any set of Majorana operators can be transformed to another set of Majorana operators by the orthogonal transformation, therefore the Hamiltonian can be always rewritten in another basis,
\begin{eqnarray}
    H = \mathrm i \sum_{ij} M_{ij} \gamma_i\, \gamma_j = 
    \mathrm i \sum_{ij} \sum_{nm} \tilde{\gamma}_n O^T_{ni} M_{ij} O_{jm} \tilde{\gamma}_m.
\end{eqnarray}
Operators $\tilde{\gamma}_n$ are Majorana operators defined by the following equation
\begin{eqnarray}
\label{eq.majorana_distribution}
    \gamma_i = \sum_n O_{in} \tilde{\gamma}_n,
\end{eqnarray}
where matrix $O$ describes the orthogonal transformation ($O^T\! O=1$) from one Majorana fermion basis to the other.
If under some transformation $O$, certain operators $\tilde{\gamma}_n$ do not appear in the Hamiltonian, then the latter operators represent MZMs. 
We would like to formalize this condition in terms of matrix $M$.

The Hamiltonian can be rewritten in terms of the new matrix $h_{nm} = \sum_{ij} O^T_{ni} M_{ij} O_{jm} $,
\begin{eqnarray}
      H = \mathrm i \sum_{n,m} \tilde{\gamma}_n h_{nm} \tilde{\gamma}_m = \mathrm i \sum_{n<m} \tilde{\gamma}_n \left(h_{nm}-h_{mn} \right)\tilde{\gamma}_m.
\end{eqnarray}
Thus, the operator $\tilde{\gamma}_n$ is a MZM operator $\Gamma_n=\tilde\gamma_n$, if $\forall_m \left(h_{nm}-h_{mn} \right)=0$.  Such operator is the integral of motion. It  commutes with the Hamiltonian $\left[H, \Gamma_n\right]=0$ simply because it does not appear in the Hamiltonian. This condition can be written as a single compact equation,
\begin{eqnarray}
\label{eq.Majorana_condition}
    \sum_m \left(h_{nm}-h_{mn} \right)^2 = 0,
\end{eqnarray}
and can be further expressed using the matrix elements of the original  Hamiltonian (see Appendix~\ref{app.details.derviation} for more details)
\begin{eqnarray}
\label{eq.Majorana_matrix_condition}
\sum_{ij} O^T_{ni} \left[(M-M^T)^2\right]_{ij} O_{jn} = 0.
\end{eqnarray}
The latter equation is similar to the eigendecomposition of the matrix $ -(M-M^T)^2 $, where the minus sign is added to obtain positive spectrum of eigenvalues,
\begin{eqnarray}
\sum_{ij} O^T_{ni} \left[-(M-M^T)^2\right]_{ij} O_{jn} = \lambda_n.\label{eq:lambdaDefinition}
\end{eqnarray}
Thus, by the numerical diagonalization of the matrix $ -(M-M^T)^2 $, and looking for  eigenvalues $\lambda_n = 0$, one finds the number of MZMs as well as their spatial structure.

If the Majorana operators $\left\{ \gamma_i \right\}$ are defined in the base of real lattice nodes, then the spatial distribution of the MZM operator $\Gamma_n$ has the general form
\begin{equation}
    \Gamma_n = \sum_{i} \alpha_{in} \gamma_i,
    \label{eq.projectors}
\end{equation}
where the coefficients $\alpha_{in}$ have to fulfill normalization condition $\sum_i \alpha_{in}^2 = 1$.
The coefficients $\left\{ \alpha_{in} \right\}$ could be obtained from a transformation that is inverse to that introduced in the Eq.~\eqref{eq.majorana_distribution},
\begin{equation}
\alpha_{in} = O_{ni}.
\label{eq:alphaDefinition}
\end{equation}

\paragraph*{Spectrum of the system.}
Next, we establish a relationship between the defined above  
eigenproblem (\ref{eq:lambdaDefinition})  and the spectrum of the  fermionic tight-binding Hamiltonian.
We express the Majorana operators in some local spinless basis in the form of a column vector $\hat{\gamma}_i =  \left(\gamma_{i}^+,\gamma_{i}^- \right)^\mathsf{T} $. 
This construction could be simply generalized to account also for systems containing spin and/or orbital degrees of freedom. 
The Majorana operators can be transformed to the standard fermionic operators written in the vector form $\hat{a}_i = \left( a_{i},a_{i}^\dagger\right)^\mathsf{T}$,
\begin{equation}
   \frac{1} {\sqrt{2}} U \hat{\gamma}_i = \hat{a}_i,
\end{equation}
where the unitary matrix $U  = \frac{1} {\sqrt{2}} \left(
\begin{array}{cc}
1 & - \mathrm i \\ 
1 &  \mathrm i \\ 
\end{array} \right) $. 
Then, the Hamiltonian (\ref{eq:general_Majorana_hamiltonian}), originally written down  in terms of Majorana fermions, can be 
expressed via standard  creation  and annihilation operators, 
\begin{equation}
    H = \mathrm i \sum_{ij} \hat{\gamma}^\mathsf{T}_i M_{ij} \hat{\gamma}_j = \sum_{ij} \hat{a}^\dagger_i \mathcal{H}_{ij} \hat{a}_j,
\end{equation}
where $\mathcal{H}_{ij} = \mathrm i U( M_{ij} - M^\mathsf{T}_{ji} ) U^{-1}$. 
In the above transformation, the non-uniqueness of matrix $M$, described in Sec.~\ref{app.details.M_matrix}, is clearly visible. 
Since the matrix $U$ is unitary, the eigenvalues of the matrix $-\left( M - M^\mathsf{T} \right)^2$ from the Eq.~(\ref{eq:lambdaDefinition}) are equal to the squares of the single-particle energies.
Thus, the spectrum of the real physical system is correctly reproduced by the spectrum of $\lambda_n$.

\begin{figure}[!b]
\centering
\includegraphics[width=\columnwidth]{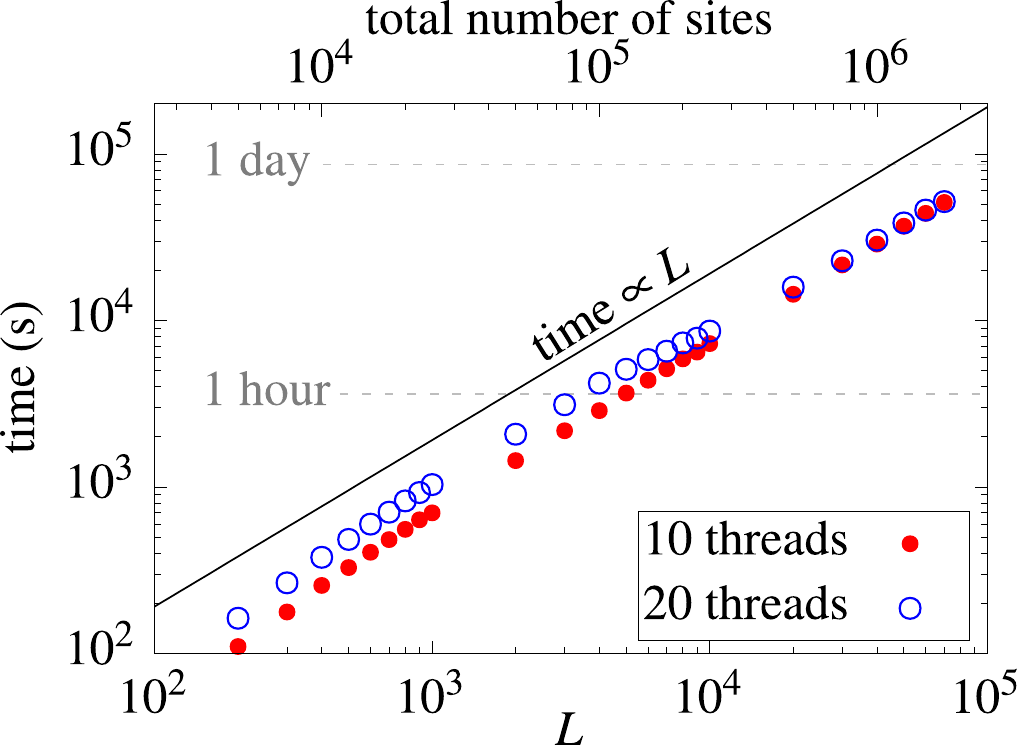}
\caption{
The calculation time versus length $L$ (or total number of sites) of the 3D Rashba nanowire (more details about model in Sec.~\ref{sec.3d_rashba}), i.e., an example system with $5\times5\times L$ sites.
The evaluation was performed on machine with Intel(R) Core(TM) i9-7900X CPU @ 3.30GHz.
Red filled and blue empty points correspond to the result obtained on 10 and 20 threads calculations, respectively.
}
\label{fig:cpu}
\end{figure}

\paragraph*{Numerical implementation.}%
The study of the topological phase in a general tight-binding model (e.g. in the presence of the disorder, untypical shape, etc.) is typically associated with the formulation of the problem in terms of the real space Bogoliubov--de~Gennes (BdG) Hamiltonian (cf.~Appendix~\ref{sec.bdg_details}). 
In practice, this is equivalent to the eigenproblem of the large sparse matrix. 
This type of problem can be solved in several methods.
The first and the most basic solution is the exact diagonalization of the Hamiltonian in the matrix form.
This method can be extended to the iterative solution of the BdG Hamiltonian based on the kernel polynomial method~\cite{weisse.wellein.06}, typically on the Chebyshev basis, called {\it the Chebyshev--Bogoliubov--de Gennes technique}~\cite{covaci.peeters.10}.
It is also worth to mention the {\sc kwant} package~\cite{groth.wimmer.14}, which can be used for studying of the topological phase within the scattering $S$ matrix method allowing the study of the transport properties of the system via calculation of the differential conductance~\cite{akhmerov.dahlhaus.11,fulga.hassler.11,rosdahl.vuik.18}. 
Additionally, this package can be treated as a tool for the construction of the real space BdG Hamiltonian.

We notice that the full diagonalization is associated with the computational complexity $\mathcal{O}(N^{3})$, while the presented method based on the Krylov subspace has the complexity $\mathcal{O}(N)$, where $N$ is the size of the Hamiltonian in the matrix form~\cite{nagai.20}.
From this, the presented method is more effective than other methods.

We implemented the described method using the \texttt{\normalsize C++} programming language and the Armadillo~\cite{sanderson.curtin.2016} library. The eigenproblem of the sparse matrix was solved by the implicitly restarted Arnoldi method implemented in the ARPACK package.
The numerical implementation is available at the public online repository \footnote{Visit the Majoranapp's Github public repository: \href{https://github.com/andywiecko/Majoranapp}{https://github.com/andywiecko/Majoranapp}}.
The program can be easily adjusted for arbitrary Hamiltonians or geometries of the system.

To evaluate our numerical implementation, we perform a 
numerical test for the 3D system (described in more details in Sec.~\ref{sec.3d_rashba}).
In this case, the number of elements in the Hamiltonian in matrix form is the biggest, i.e. the eigenproblem is computationally most expensive. 
Numerical calculations were executed on the machines with the following hardware specification Intel(R) Core(TM) i9-7900X CPU @ 3.30GHz (20~threads/10~cores).
The test of our approach, was executed in the two cases: {\it (i)} 10~threads and {\it (ii)} 20~threads (and 10~cores in the both cases).
The calculation time is shown in Fig.~\ref{fig:cpu} versus the system size and  observe nearly linear dependence.
Tiny deviations from the perfect linearity observed for large $L$ can be associated with memory storage and/or hardware bottleneck.

%%%%%%%%%%%%%%%%%%%%%%%%%%%%%%%%%%%%%%%%%%%%%
%%%%%%%%%%%%%%%%%%%%%%%%%%%%%%%%%%%%%%%%%%%%%
%%%%%%%%%%%%%%%%%%%%%%%%%%%%%%%%%%%%%%%%%%%%%

\section{Rashba chain example as a test case}
\label{sec.basic_example}

First, we start with a  test-case and study the 1D Rashba chain. 
This system is described by the Hamiltonian
\begin{eqnarray}
\mathcal{H}_\text{1D} = H_{\text{0}} + H_{\text{mag}} + H_{\text{SO}} + H_{\text{SC}} .
\end{eqnarray}
The first term describes free electrons in the chain
\begin{eqnarray}
\label{eq.kin} H_{\text{0}} &=&
      \sum_{ij\sigma} \left[ t_{ij} + \mu \delta_{ij} \right] a_{i\sigma}^\dagger a_{j\sigma} ,
\end{eqnarray}
where $a_{i\sigma}^\dagger$ ($a_{i\sigma}$) denotes creation (anihilation) operator of particle in $i$-th site with spin $\sigma\in\{\uparrow,\downarrow\}$, 
$t_{ij}$ denotes a hopping integral, while $\mu$ is the chemical potential.
We assume equal hopping between the nearest-neighbor (NN) sites (i.e., $t_{\langle i,j\rangle} = t = 1$).
The second term
\begin{eqnarray}
\label{eq.mag} H_{\text{mag}} &=& h \sum_{i\sigma\sigma'}a_{i\sigma}^\dagger \sigma^z_{\sigma\sigma'} a_{i\sigma'}  ,
\end{eqnarray}
describes the Zeeman magnetic field $h$.
Third term describes the Rashba spin--orbit coupling
\begin{eqnarray}
 \label{eq.soc_1d} H_{\text{SO}} &=&  \alpha  \sum_{i\sigma\sigma'}
 \left(
 a_{j\sigma}^\dagger \text{i} \sigma^{y}_{\sigma\sigma'} a_{i+\hat{x},\sigma'} + \mathrm{H.c.}
 \right)
 ,
\end{eqnarray}
where $\alpha$ is the spin--orbit coupling, $\hat{x}$ is the unit vector in $x$-direction (along chain), while $\sigma^{a}$ ($a = x,y,z$) denote the Pauli matrices.
The third term describes on-site superconductivity
\begin{eqnarray}
 \label{eq.sc_swave} H_{\text{SC}} &=& \Delta \sum_{i} \left( a_{i\uparrow}^\dagger a_{i\downarrow}^\dagger + \mathrm{H.c.} \right) , 
\end{eqnarray}
where $\Delta$ corresponds to the superconducting gap.

\subsection{Results}

Next,  we compare numerical results obtained from the method described in Sec.~\ref{sec.method} and from the standard technique based on the BdG equations briefly summarized in the Appendix~\ref{sec.bdg_details}.
In order to apply our algorithm for studying topological phase, first, the Hamiltonian $\mathcal{H}_{1\text D}$ should be expressed in the Majorana operator basis:
\begin{eqnarray}
\label{eq.majorana_tr}
\gamma_{i\sigma}^+ = a_{i\sigma}+a_{i\sigma}^\dagger, \quad \gamma_{i\sigma}^- = \text i(a_{i\sigma}-a_{i\sigma}^\dagger) .
\end{eqnarray}
Then, the terms of $\mathcal{H}_{1\text D}$ are given by:
\begin{eqnarray}
 \nonumber H_\text{0}+H_\text{mag} &=& -\tfrac{\text i}{2} \sum_{ij\sigma} \left[ t_{ij} + \left( \mu - \sigma h \right) \delta_{ij} \right] \left(
      \gamma_{i\sigma}^+ \gamma_{j\sigma}^- + \gamma_{j\sigma}^+ \gamma_{i\sigma}^-
      \right) , \\ && \\
%%%%%%%%%%%%%%%
\nonumber H_{\text{SO}} &=&
       - \tfrac{\text i\alpha}{2} \sum_{\langle i,j\rangle}
      \left( \gamma_{i\uparrow}^+ \gamma_{j\downarrow}^-
      + \gamma_{j\downarrow}^+ \gamma_{i\uparrow}^-
      - \gamma_{i\downarrow}^+ \gamma_{j\uparrow}^-
      - \gamma_{j\uparrow}^+ \gamma_{i\downarrow}^-
      \right) , \\ && \\
%%%%%%%%%%%%%%%
     H_{\text{SC}} &=&
     \tfrac{\text i \Delta}{2} \sum_{i}
      \left( \gamma_{i\uparrow}^+\gamma_{i\downarrow}^-
      -\gamma_{i\downarrow}^+ \gamma_{i\uparrow}^-
      \right) .
\end{eqnarray}
Here, one can notice, that a similar strategy can be used also 
for two- and three-dimensional systems, as discussed further on.
Moreover, this method can also be extended to inhomogeneous systems, e.g., with site-dependent parameters~\cite{mcginley.knolle.17,maska.gorczycagoraj.17,wieckowski.mierzejewski.20}.

\begin{figure}[!htp]
    \centering
    \includegraphics[width=\linewidth]{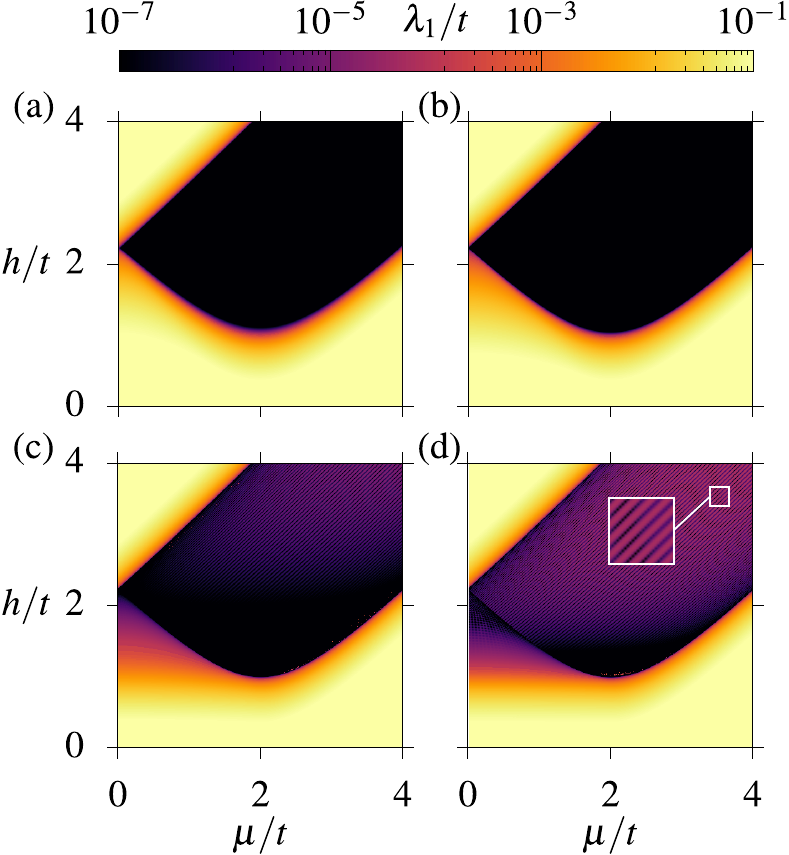}
\caption{
    The smallest eigenvalue $\lambda_1$ from Eq.~\eqref{eq:lambdaDefinition} in the case of the Rashba chain with $L=100$ sites and $\Delta/t=1$, as a function of chemical potential $\mu$ and magnetic field $h$ for different spin--orbit coupling $\alpha/t$: (a) $1.00$, (b) $0.50$, (c) $0.10$, and (d) $0.05$.
    \label{fig.lambda1}
}
\end{figure}

\begin{figure}[!hbp]
\centering
\includegraphics[width=\linewidth]{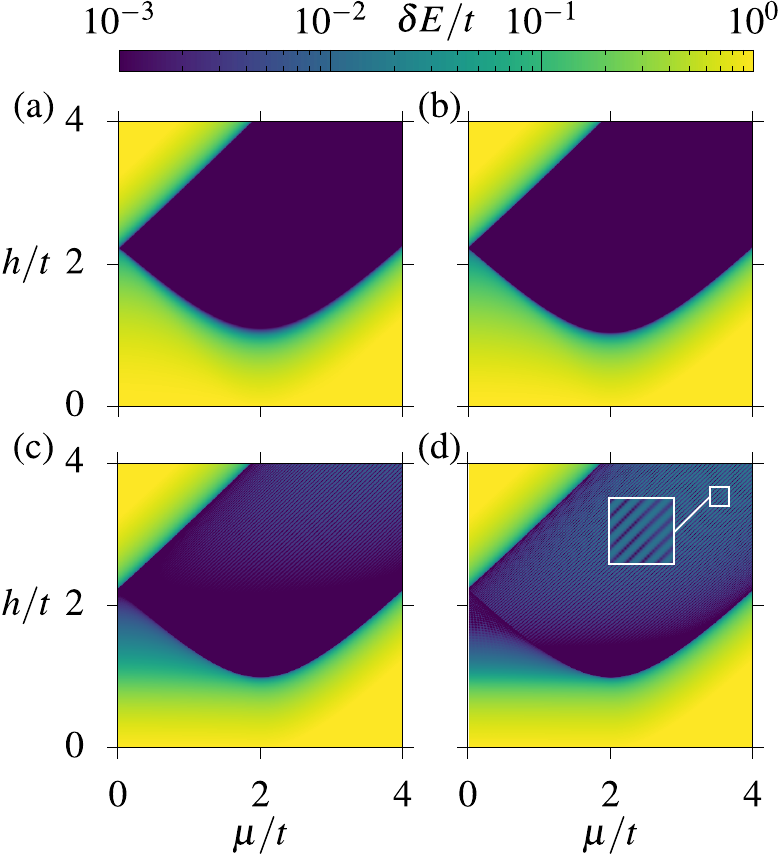}
\caption{
    The effective gap $\delta E$ for the Rashba chain with $L=100$ sites and $\Delta/t=1$, as a function of chemical potential $\mu$ and magnetic field $h$ for different spin--orbit coupling $\alpha/t$: (a) $1.00$, (b) $0.50$, (c) $0.10$, and (d) $0.05$.
    Results obtained from the Bogoliubov--de~Gennes equations technique (cf.~Fig.~\ref{fig.lambda1}).
}
\label{fig.1d_gaps}
\end{figure}

First, we study the topological phase diagram of the 1D Rashba chain.
In general, the non-trivial topological phase exists when~\cite{sato.fujimoto.09,sato.takahashi.09,sato.takahashi.10}
\begin{eqnarray}
    ( W \pm \mu )^2 + \Delta^2 < h^2 ,
    \label{eq.topo_limit}
\end{eqnarray}
where $W$ is the half-band-width, which for 1D chain $W = 2 t$.
As a consequence, Eq.~\eqref{eq.topo_limit} leads to parabolic-like boundaries of the topologically non-trivial phase~\cite{zhang.nori.16}.
The boundary between trivial and non-trivial topological phases is well visible in Fig.~\ref{fig.lambda1} and Fig.~\ref{fig.1d_gaps}.
Fig.~\ref{fig.lambda1} shows $\lambda_{1}$, i.e. the smallest eigenvalue in Eq.~(\ref{eq:lambdaDefinition}), while  Fig.~\ref{fig.1d_gaps} shows the energy gap $\delta E$ obtained within the BdG formalism. The latter quantity is defined as the difference between energies of two levels which are the closest to the Fermi energy.
In both cases, exponentially small values of parameters ($\lambda_{1}$ and $\delta E$) mark a range of parameters where a non-trivial topological phase exists.

\begin{figure}[!t]
    \centering
    \includegraphics[width=\linewidth]{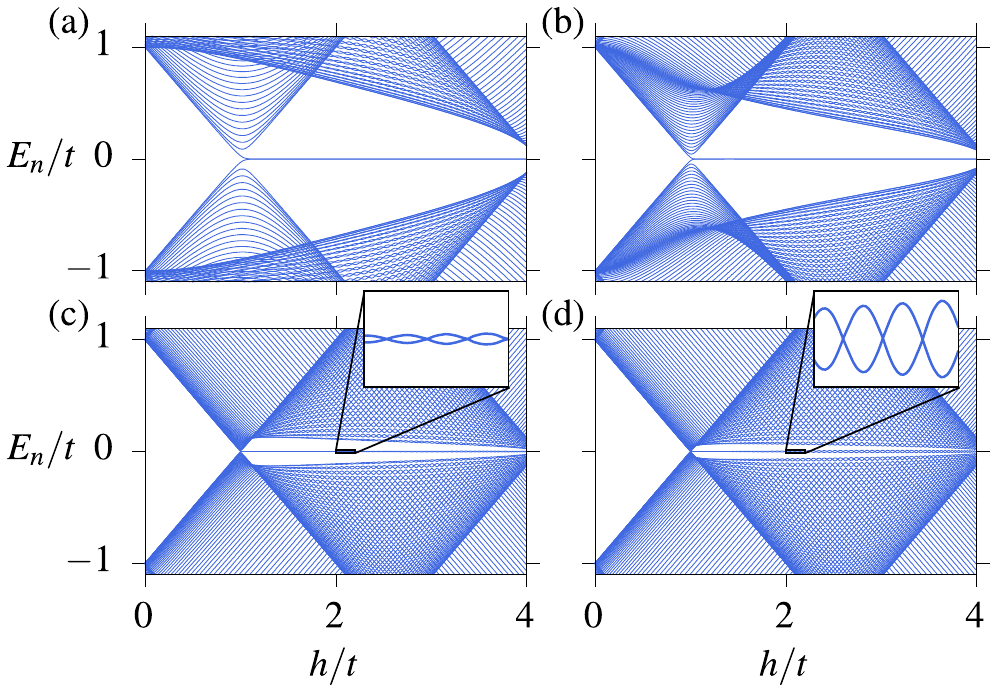}
\caption{
    The spectrum of the Rashba chain as a function of magnetic field $h$ and different value of the spin--orbit coupling $\alpha/t$: (a) $1.00$, (b) $0.50$, (c) $0.10$, and (d) $0.05$.
    Results obtained from the Bogoliubov--de~Gennes equations technique, for $L = 100$ sites,  $\mu/t = 2$, and $\Delta/t = 1$.
    Insets show oscillations and crossing of the Fermi level by the nearly-zero energy states.
}
\label{fig.warw}
\end{figure}

The emergence of the non-trivial topological phase can be discussed via the spectrum of the system (Fig.~\ref{fig.warw})~\cite{kiczek.ptok.17,kobialka.ptok.19}.
Increasing the magnetic field $h$ leads to the closing of the trivial superconducting gap.
At the some magnetic field $h_{c}$ (which for chosen parameters $h_{c} = \Delta$) topological phase transition occurs.
Further increasing $h$ leads to the reopening of the non-trivial topological gap.
The following gap strongly depends on the system parameters, e.g., spin--orbit coupling [cf. Fig.~\ref{fig.warw} panels from (a) to (d)].
In the case of finite size systems, the pair of zero (or \textit{almost} zero) energy in-gap MZMs can be found.

\begin{figure}[!b]
    \centering
    \includegraphics[width=\linewidth]{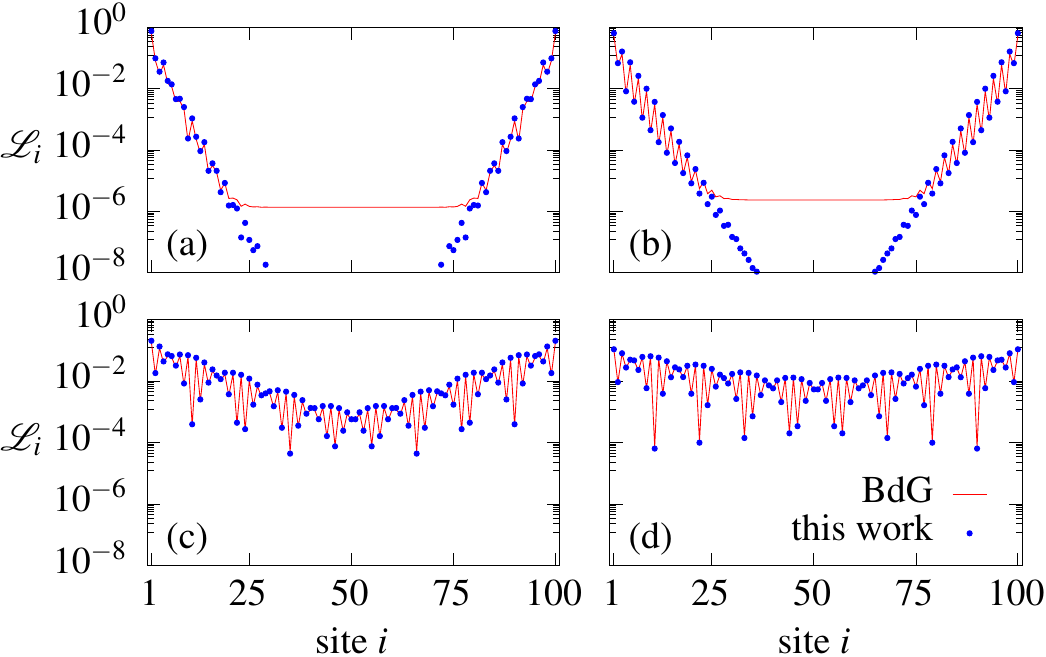}
    % h/t=1.98727
\caption{
    The comparison of Majorana modes spatial structures in the case of the Rashba chain: (normalized) local denisty of states (LDOS) shown by red line is obtained from the Bogoliubov--de~Gennes equations technique, while blue dots shown results for $\Gamma_1$ and $\Gamma_2$ obtained by the presented algorithm.
    Results for different values of the spin--orbit coupling $\alpha/t$: (a) $1.00$, (b) $0.50$, (c) $0.10$, and (d) $0.05$, in the case of chain with $L = 100$ sites and fixed $\mu/t = 2$ and $h/t \approx 2$ -- which correspond to parameters where MZMs are realized (cf. Fig.~\ref{fig.lambda1} and Fig.~\ref{fig.1d_gaps}).
}
\label{fig.ldos}
\end{figure}

The presented method allows for studying the spatial structure of the MZMs [Eq.~\eqref{eq:lambdaDefinition}]:
\begin{eqnarray}
\mathcal{L}_{i} = \sum_n \mathcal{L}_{in} ,\quad \mathcal{L}_{in} = \sum_{m=1}^{\cal M} | \alpha_{in}^{m} |^{2} ,
\label{eq.spatial}
\end{eqnarray}
where $\mathcal{L}_{i}$  describes how much $\gamma$'s located at site $i$  contribute to all  MZMs in the system. 
Therefore, $\cal M$ in Eq. (\ref{eq.spatial}), enumerates independent Majorana operators related to the single site $i$ of the system, and $n$ enumerates independent MZMs in the system.
For example, in the case of the spinless fermionic operators ${\cal M}=2$ (i.e., $\gamma_i^+$ and $\gamma_i^-$), while in the case of spinfull ones ${\cal M}=4$ (i.e., $\gamma_{i\uparrow}^+$, $\gamma_{i\uparrow}^-$, $\gamma_{i\downarrow}^+$, and $\gamma_{i\downarrow}^-$).
Coefficients $\alpha_{in}^{m}$ denote projections of the MZM operators $\Gamma_n$ on the intersite Majorana fermions $\gamma_{i}^{m}$~\cite{wieckowski.ptok.19}, cf. Eq.~(\ref{eq.projectors}).
This quantity, $\mathcal L_i$, may be compared to the zero-energy local density of states (LDOS) $\rho_{i} (\omega = 0)$, which can be obtained from the BdG technique [cf.~Eq.~(\ref{eq.bdg_ldos})].
The comparison of both quantities for the studied Rashba chain is presented in Fig.~\ref{fig.ldos}.

In a realistic (finite) system, the MZMs are characterized by exponential suppression of energy splitting \mbox{i.e.,~$[ \Gamma , \mathcal{H}_{0} ] \propto \exp ( - L / \xi_{M} )$}~\cite{dassarma.freedman.15,rainis.trifunovic.13,klinovaja.loss.12}, where $L$ is the system size.
This behavior is related to a length-scale, $\xi_{M}$, that strongly depends on the model parameters~\cite{klinovaja.loss.12,rainis.trifunovic.13,peng.pientka.15} and has a strong impact on various properties of a finite system.
%%%%%
On the one hand, the exponential decay of the MZMs with distance from the end of the chain is well visible on the logarithmic scale ~\cite{chevallier.klinovaja.16,benshach.hain.15,panaranda.aguado.18,escribano.levyyeyati.18,peng.pientka.15,klinovaja.loss.12}, which is used in Fig.~\ref{fig.ldos}.
For selected parameters shown in Figs.~\ref{fig.ldos}(c) and ~\ref{fig.ldos}(d), we observe substantial overlap of MZMs located at the opposite ends of the nanowire.
On the other hand, such overlap causes energy oscillation of the Majorana modes around the Fermi level~\cite{rainis.trifunovic.13,benshach.hain.15,escribano.levyyeyati.18,panaranda.aguado.18} shown in the insets in Fig.~\ref{fig.warw}.
%%%%%
In our calculations, the energy of  MZMs  is proportional to $\lambda_1$.
From this, exponential decay of $\lambda_1$ should be also observed.
Indeed, this is well visible in Fig.~\ref{fig.lambda1} and Fig.~\ref{fig.1d_gaps}, presenting values of $\lambda_1$ and $\delta E$, respectively.
In both cases, for a relatively small value of the spin--orbit coupling, strong oscillations of these quantities show up as a fine pattern composed of diagonal lines [cf. insets in Fig.~\ref{fig.lambda1}(d) and Fig.~\ref{fig.1d_gaps}(d)].
Moreover, the boundary of the non-trivial topological phase is rather blurred in such cases.
In this sense, the ratio $L/\xi_M$ determines how much $\lambda_1$ and $\delta E$ differ from zero.
As we mentioned above, the ratio $L/\xi_M$ can depend on many external and internal factors, e.g., the spin--orbit coupling $\alpha$ or magnetic field $h$.
Summarizing, the $\lambda_1$ correctly describes the overlap of MZMs as well as all physical consequences  which originate from such overlap.

%%%%%%%%%%%%%%%%%%%%%%%%%%%%%%%%%%%%%%%%%%%%%
%%%%%%%%%%%%%%%%%%%%%%%%%%%%%%%%%%%%%%%%%%%%%
%%%%%%%%%%%%%%%%%%%%%%%%%%%%%%%%%%%%%%%%%%%%%

\section{Two-dimensional and three-dimensional models}
\label{sec.adv_example}

In this section, we discuss results obtained from our method for a few important models hosting MZMs:
two-dimensional Kitaev model (Sec.~\ref{sec.2d_kitaev}),
two-dimensional Rashba stripe (Sec.~\ref{sec.2d_rashba}), and
three-dimensional Rashba wire (Sec.~\ref{sec.3d_rashba}).

%%%%%%%%%%%%%%%%%%%%%%%%%%%%%%%%%%%%%%%%%%%%%%
%%%%%%%%%%%%%%%%%%%%%%%%%%%%%%%%%%%%%%%%%%%%%%
%%%%%%%%%%%%%%%%%%%%%%%%%%%%%%%%%%%%%%%%%%%%%%

\begin{figure}[!t]
    \centering
    \includegraphics[width=\columnwidth]{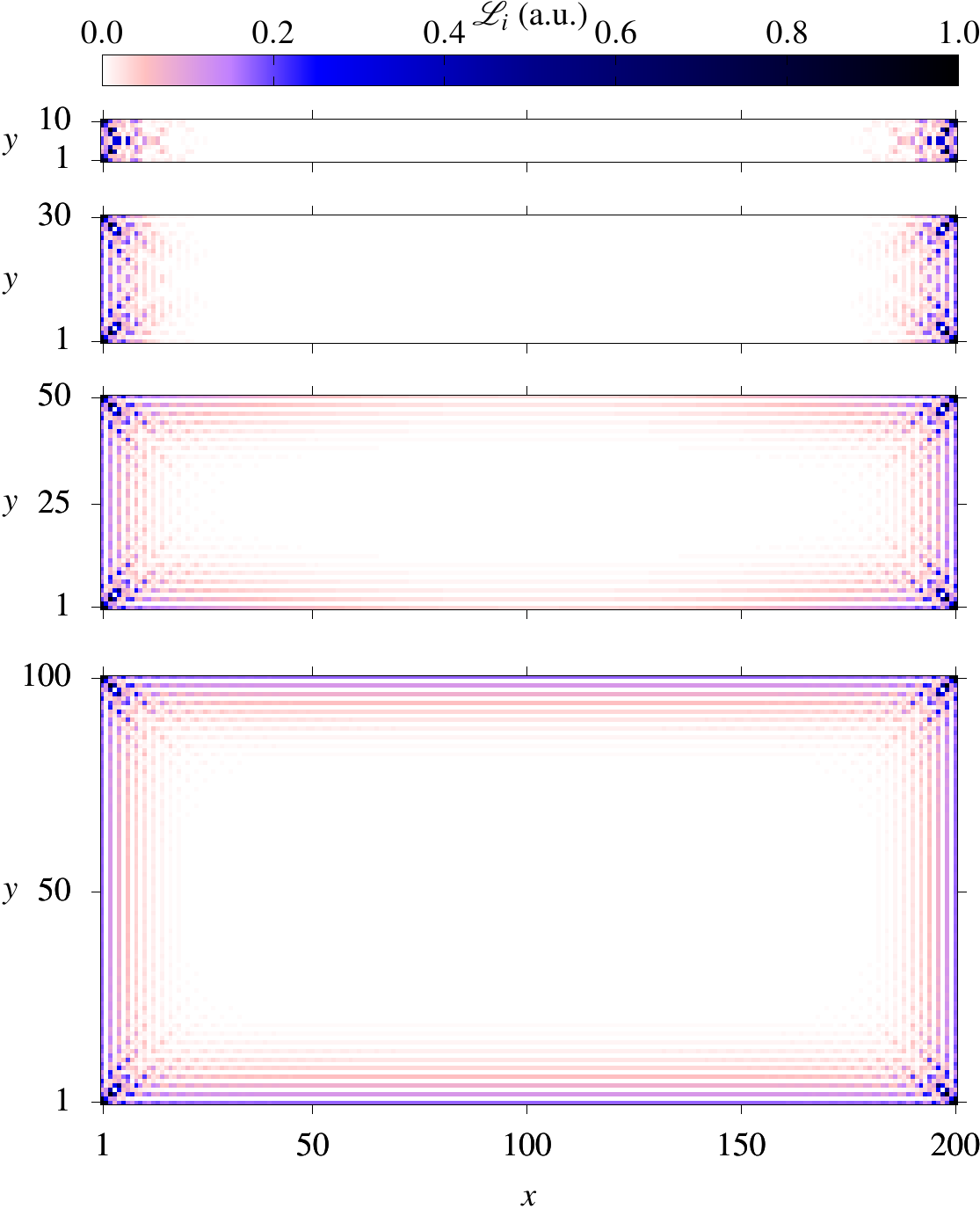}
\caption{
    The spatial structure of two MZMs under crossover from the quasi-one-dimensional to two-dimensional system (panels from top to bottom), in the Kitaev model.
    The systems have a fixed length $N_x=200$, while height $N_y$ increasing ($10$ , $30$, $50$, and $100$, for panels from top to bottom, respectively).
    The results obtained for $\Delta/t = -0.1$, and $\mu/t = - 2.0$.
}
\label{fig:kitaev2d.LDOS}
\end{figure}

\subsection{2D Kitaev model}
\label{sec.2d_kitaev}

One of the most important models studied in the context of MZMs is the Kitaev model~\cite{kitaev.01}, which describes the spinless fermions with inter-site paring in the one-dimensional chain.
This model can be extended to describe also 2D systems
\begin{eqnarray}
    \mathcal{H}_{2\text D}^{K} &=& t\sum_{\langle i,j\rangle} \left( a_i^\dagger a_j + \text{H.c.} \right) + \mu\sum_i a_i^\dagger a_i  \\
    \nonumber &+& \Delta\sum_i\left( a_i^\dagger a_{i+\hat x}^\dagger + \text i\, a_i^\dagger a_{i+\hat y}^\dagger + \text{H.c.}\right) .
\end{eqnarray}
Here, operators $a^{\dagger}_{i}$ and $a_{i}$ denote creation and annihilation of the spinless fermion in the \textit{i}-th site of the lattice.
The first term describes hopping between the neighboring sites. 
The last term corresponds to the $p+\text ip$ pairing, which can be effectively induced in some systems by the interplay between the \textit{s}-wave superconductivity, the spin--orbit coupling and the external magnetic field~\cite{gorkov.rashba.01,zhang.tewari.08,ptok.rodriguez.18}.

Our results can be compared with the results shown in Ref.~\cite{potter.lee.10}.
First, we accurately reproduce the crossover from the quasi-one-dimensional ($N_x \gg N_y$) to exact 2D ($N_x \simeq N_y$) system (cf. Fig. 1 in Ref.~\cite{potter.lee.10} and Fig.~\ref{fig:kitaev2d.LDOS} in this work).
In our case, the spatial structure can be found from Eq.~(\ref{eq.spatial}).
In the quasi-one-dimensional system, the MZMs are localized at the system edges (top panel at Fig.~\ref{fig:kitaev2d.LDOS}).
Modification of the $N_y / N_x$ ratio leads to crossover from quasi-one-dimensional system to a two-dimensional one (cf. panels from top to bottom at Fig.~\ref{fig:kitaev2d.LDOS}).
During this crossover,  the MZMs relocate from the ends of the chain-like system towards the edges of the 2D cluster.
Moreover, the exponential decay of the MZMs from edge to the center of the system is well visible on the length-scale $\xi_M \approx 10 a \propto \Delta / t$, where $a$ is the lattice constant).

\begin{figure}[!t]
    \centering
    \includegraphics[width=\columnwidth]{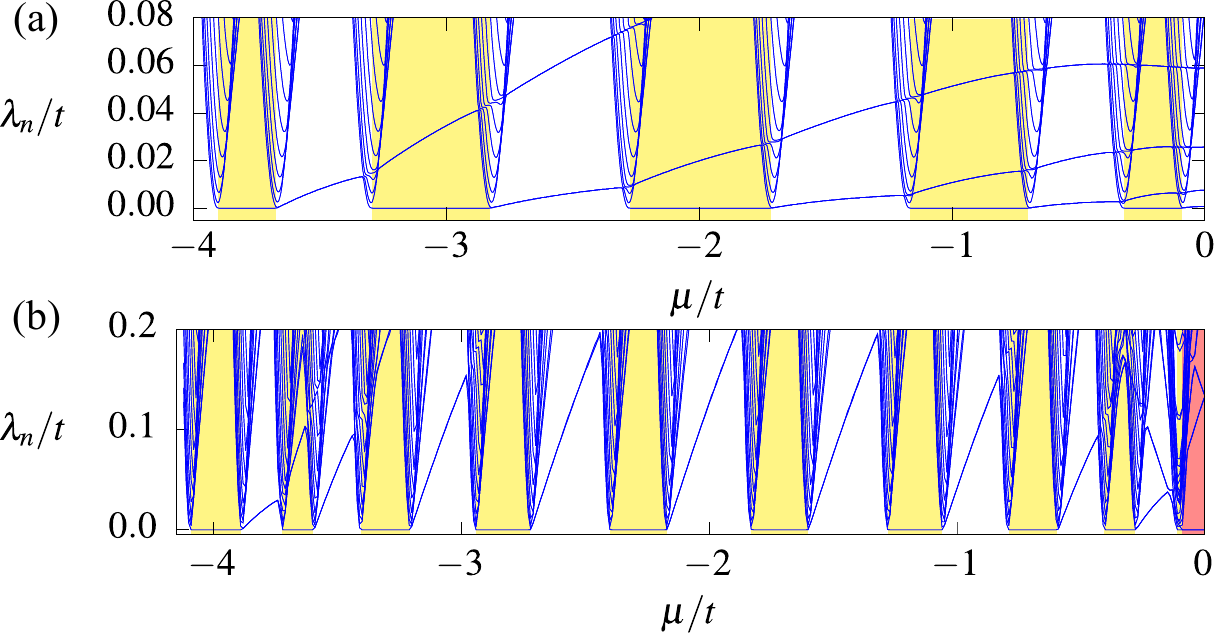}
\caption{
    The spectrum of $\lambda_n$ from Eq.~\eqref{eq:lambdaDefinition} as a function of chemical potential $\mu$.
    (a) The Results for the two-dimensional Kitaev model (for $\Delta/t=-0.1$, $N_x=100$, and $N_y=10$), and 
    (b) for two-dimensional Rashba model (for $\alpha/t=0.1$, $\Delta/t=0.1$, $h_z/t=0.2$, $N_x=150$, and $N_y=10$).
    In both cases, the yellow and red shaded areas mark range of $\mu$ for which system hosts 2 and 4 MZMs, respectively.
}
\label{fig:kitaev2d.lambda}
\end{figure}

\begin{figure}[!b]
    \centering
    \includegraphics[width=\columnwidth]{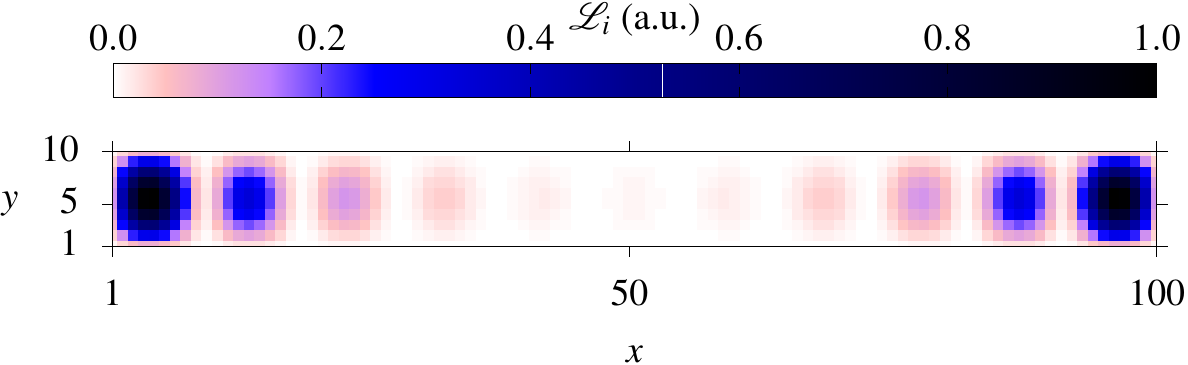} 
\caption{
    The spatial structure of two MZMs, for Rashba model with in-plane magnetic field along system.
    The results obtained for quasi-one dimensional system with $N_x \times N_y = 100 \times 10$ for $\alpha/t=0.1$, $\Delta/t=0.1$, $h_x/t=0.2$, $h_z/t=0$, $\mu/t=4$, and $\Delta/t=0.1$.
}
    \label{fig:rashba2d.LDOS1}
\end{figure}

Next, we single out a range of $\mu$ for which the system hosts MZMs.
This can be done in a relatively simple way by studying the energy spectrum and the existence of the zero energy in-gap states (cf. Fig.~2 in Ref.~\cite{potter.lee.10}).
In our case, a similar information is encoded in the spectrum $\lambda_n$ [cf. Eq.~(\ref{eq:lambdaDefinition})], which is shown in Fig.~\ref{fig:kitaev2d.lambda}(a).
As one can see, there are several windows of chemical potential (marked by yellow background color) for which there exist several solutions with $\lambda_n$ approximately equal $0$ and is separated from other states by a relatively large gap.
Moreover, because the system is \textit{quasi}-1D, in some range of $\mu$ we observed induced existence additional in-gap states associated with edge states in such system.

Here, we specify the numerical condition for the presence of the MZMs.
In an infinite system, the existence of MZMs is equivalent to the presence of eigenmodes with $\lambda_{n} = 0$ and each such eigenmode represents a single MZM.
However, the numerical studies are carried out for large but finite systems. Then, the MZMs do overlap with each other and are not strict integrals of motion. 
Hence, we look for solutions $\lambda_{n} < \epsilon$, where  we take $\epsilon = 10^{-6}$.
In order to improve identification of the MZMs and to eliminate ambiguity connected with the choice of $\epsilon$, we check whether the nearly-degenerate, eigenvalues $\lambda_n<\epsilon $  are separated from the rest of the spectrum by a gap that is much larger than $\epsilon$.
The existence of the latter gap is clearly visible in  Fig.~\ref{fig:kitaev2d.lambda} for highlighted windows of $\mu$. 
Finally, we notice that the latter gap is very important for the efficiency of the Lanczos algorithm.

%%%%%%%%%%%%%%%%%%%%%%%%%%%%%%%%%%%%%%%%%%%%%%
%%%%%%%%%%%%%%%%%%%%%%%%%%%%%%%%%%%%%%%%%%%%%%
%%%%%%%%%%%%%%%%%%%%%%%%%%%%%%%%%%%%%%%%%%%%%%

\subsection{2D Rashba stripe}
\label{sec.2d_rashba}

Similarly to the Kitaev model, also the Rashba chain described in Sec.~\ref{sec.basic_example}, can be extended to the 2D form
\begin{eqnarray}
\mathcal{H}_{2\text D}^{R} = H_{\text{0}} + H_{\text{mag}} + H_{\text{SO}} + H_{\text{SC}} .
\end{eqnarray}
The kinetic term, $H_{\text{0}}$, and the superconducting term $H_{\text{SC}}$ are unchanged and take the same form as previously [cf. Eq.~\eqref{eq.kin} and~\eqref{eq.sc_swave}, respectively].
However, we assume a more general form of the term describing the external magnetic field
\begin{eqnarray}
H_{\text{mag}} &=& h_x \sum_{i\sigma\sigma'}a_{i\sigma}^\dagger \sigma^x_{\sigma\sigma'} a_{i\sigma'} + h_z \sum_{i\sigma\sigma'}a_{i\sigma}^\dagger \sigma^z_{\sigma\sigma'} a_{i\sigma'} ,
\end{eqnarray}
where the magnetic field can be arranged in $x$ ($h_{x} \neq 0$ and $h_z = 0$) or $z$ ($h_{x} = 0$ and $h_z \neq 0$) directions.
Also the Rashba spin--orbit coupling term takes the two-dimensional generalized form
\begin{eqnarray}
\nonumber H_{\text{SO}} &=& \alpha \sum_{i\sigma\sigma'} \left[ a_{i\sigma}^\dagger \text i\sigma^x_{\sigma\sigma'} a_{i+\hat{y},\sigma'} + a_{i\sigma}^\dagger \text i\sigma^y_{\sigma\sigma'} a_{i+\hat{x},\sigma'} + \text{H.c.} \right] . \\
\end{eqnarray}

\begin{figure}[!t]
    \centering
\includegraphics[width=\columnwidth]{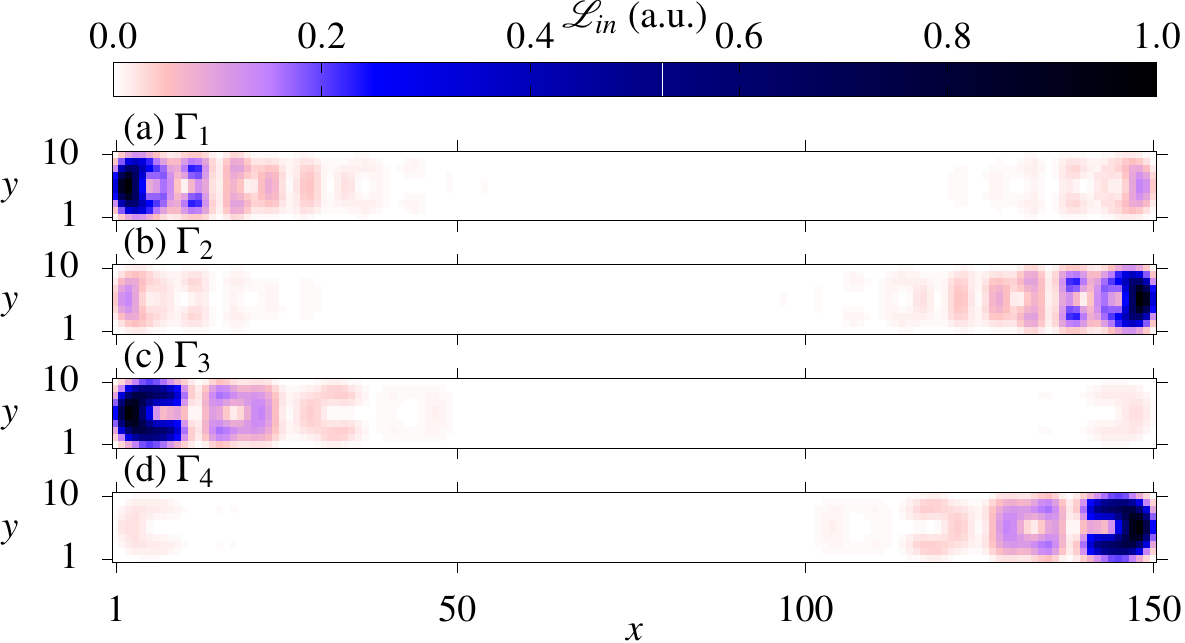}
\caption{ 
The spatial structure of MZMs for Rashba model with out-of-plain magnetic field.
    Panels from top to bottom show localization of the separate Majorana modes $\Gamma_n$ ($n = 1,2,3,4$).
    The results obtained for quasi-one dimensional system with $N_x \times N_y = 150 \times 10$ for $\alpha/t=0.1$, $\Delta/t=0.1$, $h_x/t=0$, $h_z/t=0.2$, and $\mu/t=0.05$.
}
    \label{fig:rashba2d.LDOS3}
\end{figure}

\begin{figure}[!b]
    \centering
    \includegraphics[width=\columnwidth]{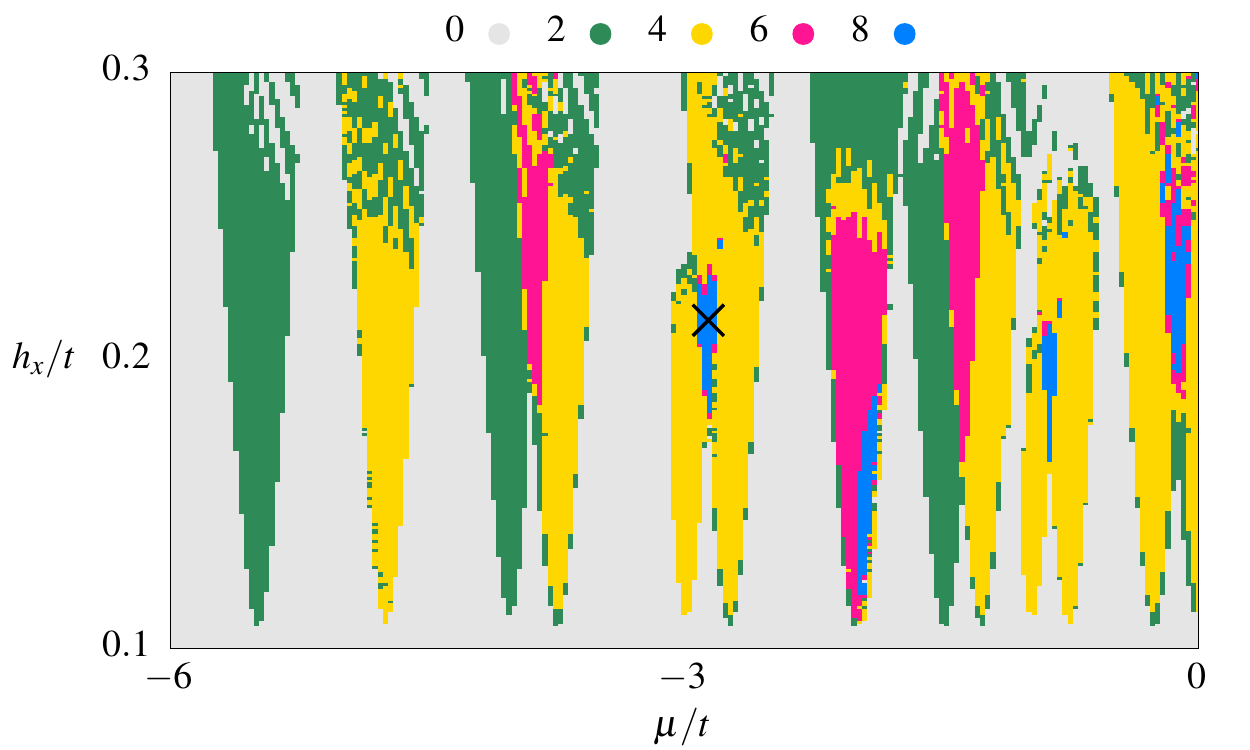}
\caption{
    The number of independent MZMs as a function of chemical potential $\mu$ and magnetic field $h_x$ for the three-dimensional Rashba model.
    The results for $\alpha/t= 0.1$, $\Delta/t=0.1$, and $N_{x} \times N_{y}\times N_z = 150 \times 5\times 5$.
    Black cross shows parameters used in Fig.~\ref{fig:rashba3d.no8}.
    }
    \label{fig:rashba3d.diagram}
\end{figure}

\begin{figure*}[!pt]
    \centering
    \includegraphics[width=\textwidth]{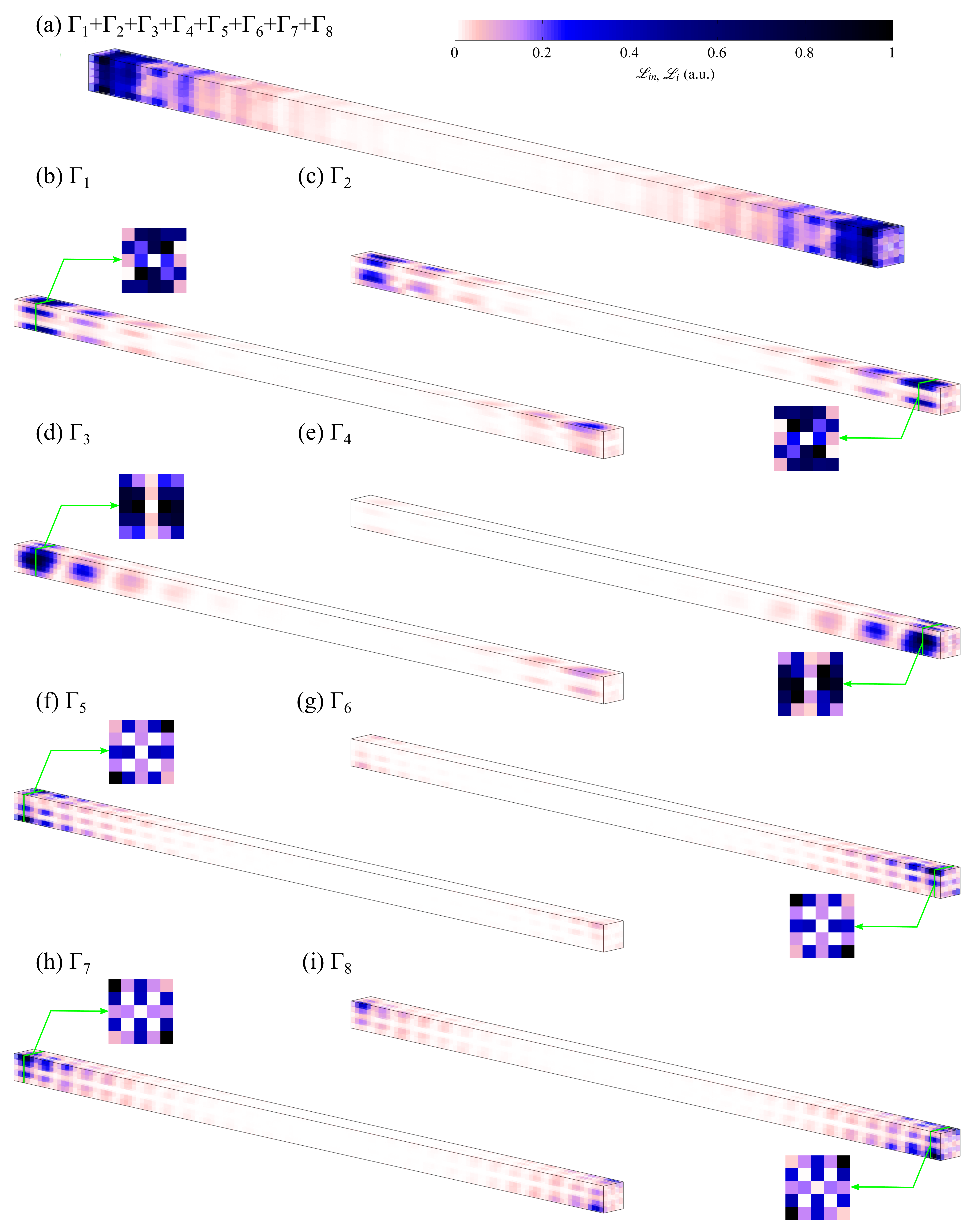}
\caption{
    The spatial structure of MZMs for three-dimensional Rashba model with magnetic field along wire.
    Panel (a) shows the spatial structure for all eight MZMs combined, while panels from (b) to (i) show localization of the separated MZMs $\Gamma_n$ ($n = 1,2,\dots,8$).
    Additional insets show the wire cross-section to the precise analyzed MZM localization.
    The results form the $N_x \times N_y \times N_z= 150 \times 5 \times 5$ for $\alpha/t = 0.1$, $\Delta/t =0.1$, $h_x/t=0.214$, and $\mu/t = -2.859$.
}\label{fig:rashba3d.no8}
\end{figure*}

\begin{figure*}[!t]
    \centering
    \includegraphics[width=\textwidth]{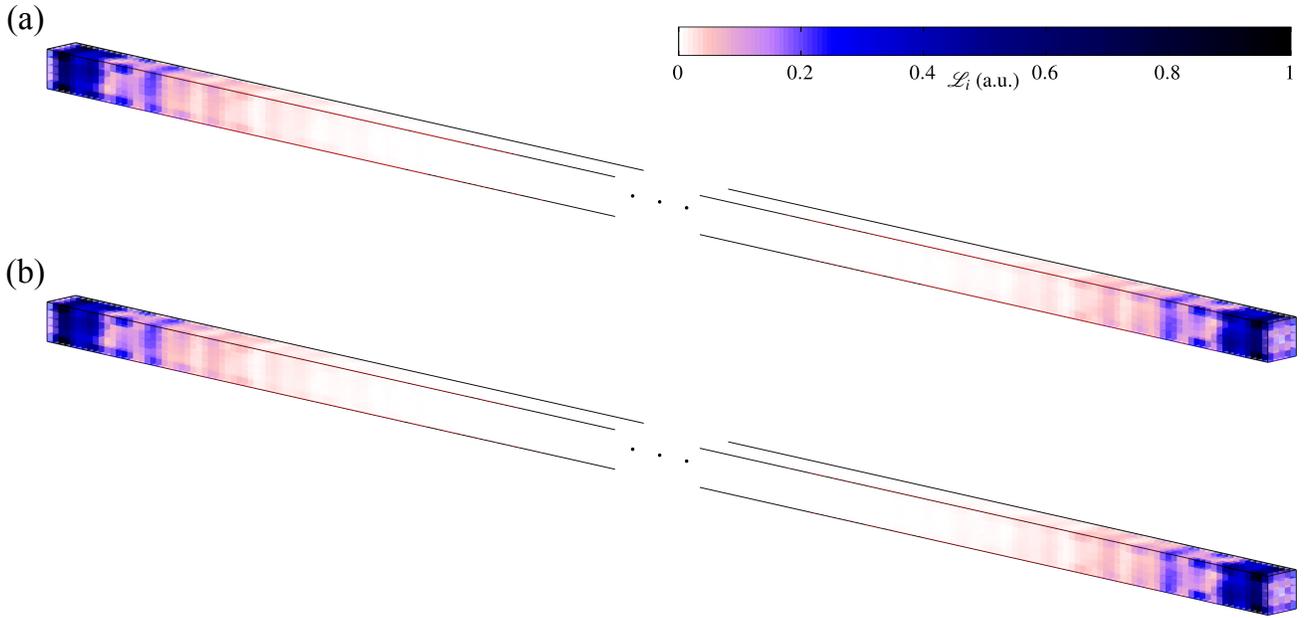}
    \caption{
    Additional results form 3D Rashba nanowire (cf.~Sec.~IV.C in main text), in the case of the system with
    (a) $N_x\times N_y\times N_z = 1\;500\times5\times5$, and
    (b) $15\;000\times5\times5$ sites.
    }\label{fig:longlongwire}
\end{figure*}

The boundaries of non-trivial phases with MZMs in this model have been determined by analysing the spectrum $\lambda_n$, which is shown in Fig.~\ref{fig:kitaev2d.lambda}(b).
The presented results reveal the multiband nature of the two-dimensional system~\cite{dumitrescu.roberts.15}.
Increasing the chemical potential $\mu$ (for fixed magnetic field $h_z$ and $h_x = 0$) we observe several windows of $\mu$ for which MZMs are hosted in the system.
This behavior is associated with the crossing of the Fermi level by the bottom of subsequent sub-bands of the quasi-two-dimensional system.
The result is in agreement with the topological phase diagram of the system discussed in Ref.~\cite{gibertini.taddei.12}.

Our method allows for study not only the topological phase diagram (i.e. the ranges of $\mu$ and $h_z$, when the system  hosts the MZMs) but also the number of independent pairs of MZMs in the system.
Due to presence of the spin--orbit coupling, the number as well as the spatial structure  of the MZMs typically depends on the direction of the magnetic field with respect to the system's plain.
For comparison, we present  results for the in-plain magnetic field   (Fig.~\ref{fig:rashba2d.LDOS1}) and the out-of-plain field (Fig.~\ref{fig:rashba2d.LDOS3}).
In the former case, we obtain  a single pair of MZMs and its spatial structure remains in agreement with  the result from the literature [cf.~Fig.~3(a) in Ref.~\cite{gibertini.taddei.12}].
It is interesting that in the case the out-of-plain magnetic field we can find four independent MZMs.
In Fig.~\ref{fig:rashba2d.LDOS3} we have shown their spatial structures, separately for each MZM.
As expected, MZMs can be found in pairs which are symmetrically placed with respect to the center of the system [cf. (a) and (b) panel or (c) and (d) panel in Fig.~\ref{fig:rashba2d.LDOS3}].
Moreover, each pair oscillates in the real-space with different periodicity (cf. two top panels and two bottom panels).
Such behavior has been reported before~\cite{livanas.sigrist.19,xie.law.20}.
Here, we must have in mind that the explicit form of MZMs $\Gamma_{n}$ is not uniquely defined.
Arbitrary orthogonal transformation $\mathcal O$ applied to a vector of MZMs ${\bm \Gamma} = ( \Gamma_1,\Gamma_2,\dots,\Gamma_{\mathcal N} )^T$ gives a rotated vector of independent and orthogonal MZMs $\widetilde{\bm \Gamma} = [\widetilde\Gamma_1,\widetilde\Gamma_2,\dots,\widetilde\Gamma_{\mathcal N}]^T$
\begin{equation}
    \widetilde{\bm\Gamma} = \mathcal{O} \bm{\Gamma}.
\end{equation}
Here, $\mathcal O$ is an orthogonal matrix, $\mathcal O ^T\!\!\mathcal O=1$, and $\mathcal N$  is the number of independent MZMs in the system.
Since rotation does not change commutation relations, $\{\widetilde \Gamma_n,\widetilde \Gamma_m\}=\{ \Gamma_n, \Gamma_m\} = 2\delta_{nm}$, or the commutation with the Hamiltonian $[H,\Gamma_n]=[H,\widetilde\Gamma_n]=0$, the transformed MZMs $\widetilde{\bm \Gamma}$ consists independent, conserved MZMs. 
Such rotation obviously does not change physical properties of the system.

%%%%%%%%%%%%%%%%%%%%%%%%%%%%%%%%%%%%%%%%%%%%%%
%%%%%%%%%%%%%%%%%%%%%%%%%%%%%%%%%%%%%%%%%%%%%%
%%%%%%%%%%%%%%%%%%%%%%%%%%%%%%%%%%%%%%%%%%%%%%

\subsection{3D model of a Rashba wire}
\label{sec.3d_rashba}

The numerical efficiency of our method allows for studying also three-dimensional systems.
Here, as an example, we investigate the extended Rashba model on a three-dimensional lattice
\begin{eqnarray}
\mathcal{H}_{3\text D}^{R} = H_{\text{0}} + H_{\text{mag}} + H_{\text{SO}} + H_{\text{SC}} .
\end{eqnarray}
The kinetic and superconducting terms are the same as previously and are expressed, respectively, by  Eqs.~\eqref{eq.kin} and~\eqref{eq.sc_swave}. We assume the magnetic field along the $x$-direction
\begin{eqnarray}
    \mathcal{H}_\text{mag} &=& h_x\sum_{i\sigma\sigma'} a_{i\sigma}^\dagger \sigma^x_{\sigma\sigma'} a_{i\sigma'}.
\end{eqnarray}
The spin--orbit coupling takes the form
\begin{eqnarray}
\nonumber H_{\text{SO}} = \alpha \sum_{i\sigma\sigma'} && \left[ 
a_{i\sigma}^\dagger \text i \sigma^x_{\sigma\sigma'} a_{i+\hat{y},\sigma'} + a_{i\sigma}^\dagger \text i \sigma^y_{\sigma\sigma'} a_{i+\hat{x},\sigma'} \right. \\
 && + \left. a_{i\sigma}^\dagger \text i \sigma^z_{\sigma\sigma'} a_{i+\hat{z},\sigma'} + \text{H.c.} \right] .
\end{eqnarray}

Similarly to the case of a two-dimensional system,  the topological phase diagram is composed of several parabolic-like structures shown in Fig.~\ref{fig:rashba3d.diagram}.
This property of the topological phase diagram has been reported previously but only for 2D models~\cite{javier.llorenc.15,sedlmayer.aguiarhualde.16,kaladzhyan.bena.17}.
When the Fermi level crosses the bottom of the lowest band, i.e. for $\mu/t \approx -5.5$, one pair of MZMs occurs in system as marked by the first green parabola in the left part of Fig.~\ref{fig:rashba3d.diagram}.
For larger $\mu$, there exist edge states containing much more independent MZMs.
In particular, one can find  up to eight independent MZMs in the studied system and the corresponding model parameters are marked in Fig.~\ref{fig:rashba3d.diagram} by a blue color.

One particular realization of eight MZMs is shown in Fig.~\ref{fig:rashba3d.no8}.
The spatial structure of MZMs shows clear similarity with respect to the previously discussed 2D case.
For instance,  all the  MZMs can be grouped in pairs which are  symmetrically distributed with respect to the geometric center of the system (cf. panels in Fig.~\ref{fig:rashba3d.no8} shown in a single row).
The latter symmetry is also well visible from the cross-sections shown in the insets.
Our analysis shows that the Majorana states are not located at the surface of the wire. 
Instead, they are distributed throughout almost the entire cross-sections of the wires. 
Additionally, the previously discussed real-space oscillations of the MZMs are visible along the wire and the periodicity  of these oscillations can be very different for various pairs of the MZMs. 
Generally, the complex spatial structure of MZMs originates from mutual orthogonality of multiple pairs of MZMs.

Increasing the length of the nanowire $N_x$ does not change the main result.
In Fig.~\ref{fig:longlongwire} one can find the results of MZMs spatial structure for system with $N_x\times N_y\times N_z = 1\;500\times5\times5$ and $15\;000\times5\times5$ sites, i.e., containing $37\;500$ and $375\;000$ sites, respectively.
This example shows, that our algorithm provides fully convergent numerical results for 3D models hosting the MZMs, which spatial structure is insensitive to the length of the nanowire.
Studying the 3D model, one observes a nontrivial variation of the spatial structure of MZM on all the edge surfaces.
The latter feature (obviously missing in the 1D models) may be studied experimentally within the scanning tunneling spectroscopy.

%%%%%%%%%%%%%%%%%%%%%%%%%%%%%%%%%%%%%%%%%%%%%
%%%%%%%%%%%%%%%%%%%%%%%%%%%%%%%%%%%%%%%%%%%%%
%%%%%%%%%%%%%%%%%%%%%%%%%%%%%%%%%%%%%%%%%%%%%

\section{Topological edge modes in magnetic nanoisland}
\label{sec.island}

Finally, we discuss the realization of the topological in-gap states in the magnetic nanoisland.
Such setup has been realized experimentally in the form of Co magnetic island grown on Si and covered by Pb monolayer~\cite{menard.guissart.17} or Fe monoatomic islands deposited on an oxygen-reconstructed Re surface~\cite{palaciomorales.mascot.19}.
In such a situation, the interplay between intrinsic spin--orbit coupling, superconductivity induced by proximity effect, and magnetic moments of the nanoisland can lead to the realization of the 2D topological domain.

Typically, nanoislands have an irregular shape and can be described by the Hamiltonian with site-dependent parameters:
\begin{eqnarray}
\mathcal{H} &=& \sum_{ij\sigma} \left[ t_{ij} + \mu \delta_{ij} \right] a_{i\sigma}^\dagger a_{j\sigma} + \sum_{i\sigma\sigma'} h_i a_{i\sigma}^\dagger \sigma^z_{\sigma\sigma'} a_{i\sigma'} \\
\nonumber &+& \alpha \sum_{i\sigma\sigma'} \left( a_{j\sigma}^\dagger \text{i} \sigma^{x}_{\sigma\sigma'} a_{i+\hat{y},\sigma'} + a_{j\sigma}^\dagger \text{i} \sigma^{y}_{\sigma\sigma'} a_{i+\hat{x},\sigma'} + \mathrm{H.c.} \right) \\
\nonumber &+& \Delta \sum_{i} \left( a_{i\uparrow}^\dagger a_{i\downarrow}^\dagger + \mathrm{H.c.} \right) ,
\end{eqnarray}
where $a_{i\sigma}^\dagger$ ($a_{i\sigma}$) denotes creation (anihilation) operator of particle in $i$-th site with spin $\sigma\in\{\uparrow,\downarrow\}$, $\hat n_{i\sigma} = a_{i\sigma}^\dagger a_{i\sigma}$,
$t_{ij}$ denotes a hopping integral, $\mu$ is the chemical potential, $h_i$ is site dependent magnetic field, $\alpha$ is the spin--orbit coupling, while $\Delta$ corresponds to the superconducting gap.
Here $h_i$ corresponds to the magnetic moments of the atoms inside the nanoisland. 
When $h_i$ is strong enough, the topological domain occurs.

\begin{figure}[!b]
\centering
\includegraphics[width=\columnwidth]{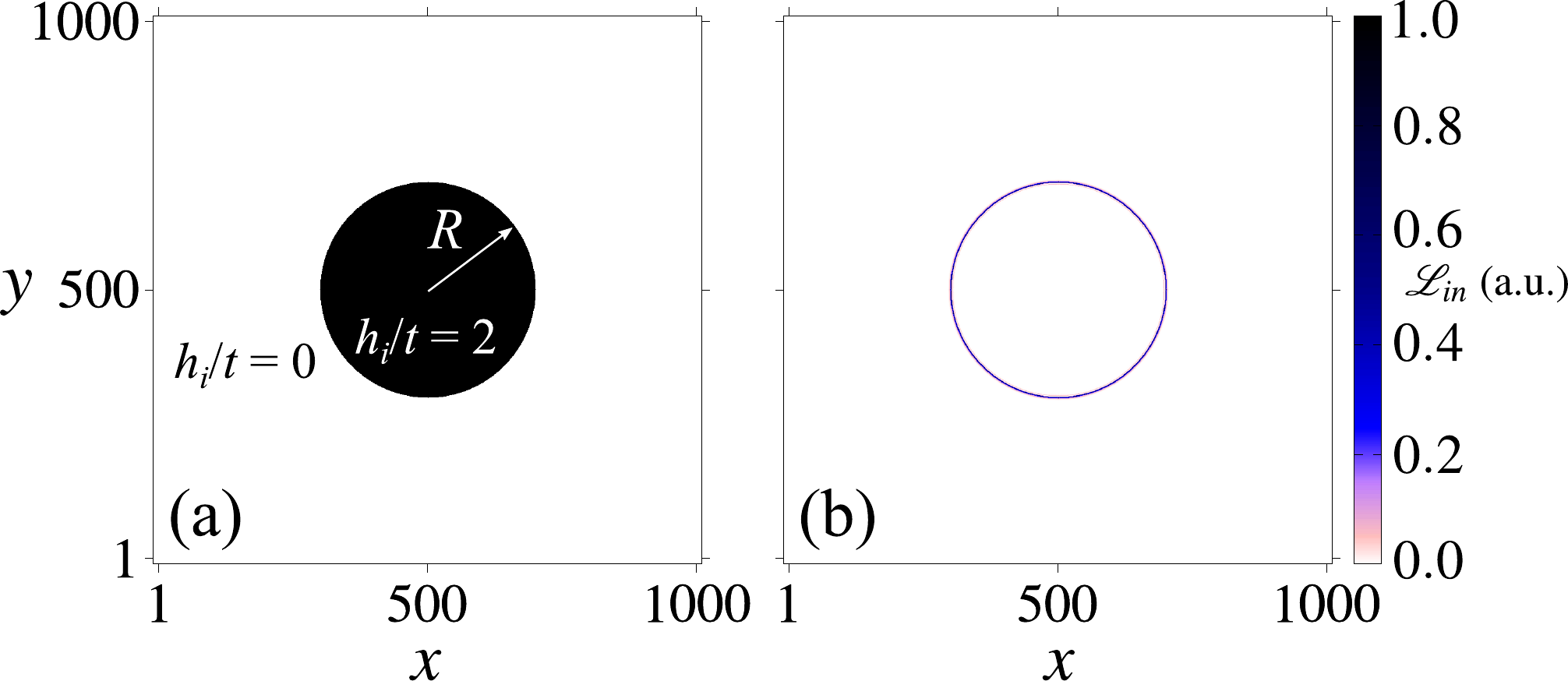}
\caption{
(a) The magnetic nanoisland in the shape of the circle of the radius $R=200$ sites (with magnetic field $h_i/t = 2$), represented by black shape, is surrounded by neutral system (without magnetic field).
(b) Localization of the MZMs ($\Gamma_n$ for $n=1,2$) around of the edge of nanoisland.
The results for $\alpha/t=1$, $\Delta/t=1$, $\mu/t=4$, $N_x\times N_y = 1\;000 \times 1\;000$.
}
\label{fig:circle1}
\end{figure}

In our calculations we take $\alpha/t=1$, $\Delta/t=1$, and $\mu/t=4$. 
Additionally, we assume $h_i/t=2$ inside nanoisland and $0$ otherwise.
Similarly to the previous cases, in the homogeneous system transition to the non-trivial topological phase occurs at some critical magnetic filed, here at $h_{c} = \sqrt{ \Delta^{2} + ( 4 t - \mu^{2} ) }$~\cite{sato.fujimoto.09,sato.takahashi.09,sato.takahashi.10}.
In the case of the inhomogeneous system, the non-trivial topological phase can be realized locally in form of the domain for which the above condition is fulfilled~\cite{liu.drummond.12,ptok.cichy.18,ptok.alspaugh.20} (i.e. for sites in which the effective magnetic field is bigger than $h_{c}$).
As we can see, for our set of parameters, the non-trivial phase is realized inside the nanoisland and surrounded by a trivial phase.
For the sake of simplicty, we take the system in the shape of the circle (Fig.~\ref{fig:circle1}). 
The MZMs are uniformly distributed along the edge independently of the details of the  geometry, as shown in Fig.~\ref{fig:circle1}(b).
Modification of the island's geometry does not change the main results presented in the Section.

In contrast to quasi-1D structures discussed in the previous sections, in the 2D case we observed several topological in-gap states associated with the the edge states around the nanoisland boundary [Fig.~\ref{fig:circle1}(b)].
This is typical behavior observed in the case of the finite-size 2D system, associated with the realization of the Majorana chiral modes~\cite{li.neupert.16}.
Here, it should be noticed that the spectrum of the system does not contain exact zero energy Majorana modes, as reported before~\cite{rontynen.ojanen.15,rachel.mascot.17,kobialka.domanski.19}.
Indeed, we have checked this behavior in the case of nanoisland with a circular shape, sketched in Fig.~\ref{fig:circle1}(a). 
The size of the system (with periodic boundary condition) is fixed and contains $1\;000\;000$ sites.

\begin{figure}[!t]
\centering
\includegraphics[width=\columnwidth]{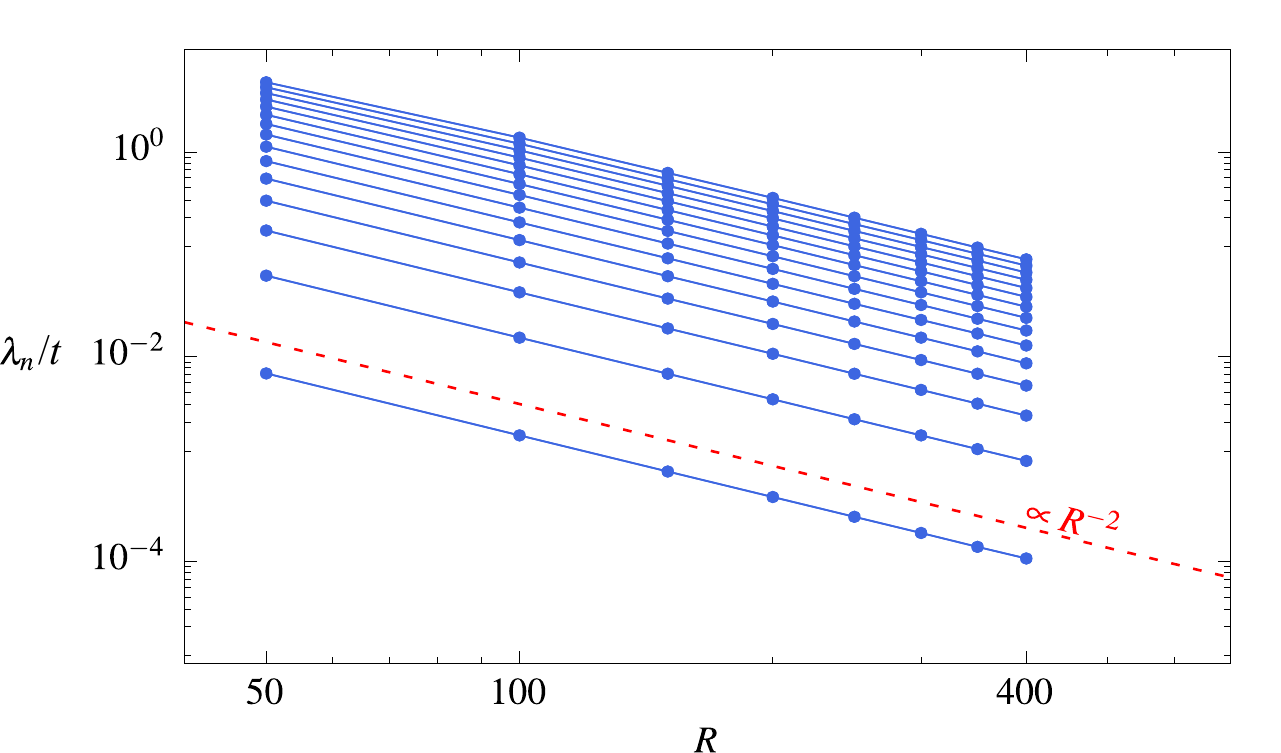}
\caption{
The spectrum of $\lambda_n$ (the lowest 30 eigenvalues) as a function of radius $R$ of the circular magnetic nanoisland (cf. Fig.~\ref{fig:circle1}).
The results for $\alpha/t=1$, $\Delta/t=1$, $\mu/t=4$, $N_x\times N_y = 1\;000 \times 1\;000$.
}
\label{fig:circle2}
\end{figure}

Increasing  the radius $R$, we check how the energy spectrum scales with the  size of the island.
In Fig.~\ref{fig:circle2} we show the spectrum $\{ \lambda_n \}$ which corresponds to the squares of the single-particle energies,
as it has been demonstrated in Sec.~\ref{sec.method}.
One observes that multiple eigenvalues decay with the system sizes as $1/R^2$, which corresponds to the decrease of the Hamiltonian spectrum as $1/R$.
In other words, eigenvalues of the Hamiltonian are inversely proportional to the edge length.
In the thermodynamic limit ($R \rightarrow \infty$), one obtains several Majorana modes with relatively small energies.
However, we observed slower decay of the smallest eigenvalues to the zero energy than in 1D systems, where the zero energy mode is observed already in  systems with $\sim 100$ sites (cf.~Sec.~\ref{sec.basic_example}).
Independently of this, the exact zero energy mode is not realized in the 2D system.

%%%%%%%%%%%%%%%%%%%%%%%%%%%%%%%%%%%%%%%%%%%%%
%%%%%%%%%%%%%%%%%%%%%%%%%%%%%%%%%%%%%%%%%%%%%
%%%%%%%%%%%%%%%%%%%%%%%%%%%%%%%%%%%%%%%%%%%%%

\section{Summary}
\label{sec.summary}

The main purpose of the present work is to derive and test a new method for identifying the MZMs in an arbitrary tight-binding model without many-body interactions or when the latter interactions may be introduced within a mean-field-like approximation. 
The basic idea behind our method is to find an orthogonal transformation of the Majorana fermions (not to be confused with the MZMs) such that selected fermions are missing in the transformed Hamiltonian. 
Then, the missing Majorana operators represent true zero modes, i.e. the MZMs.
However, as we show in Sec.~\ref{sec.method}, the presented method can be generalized to the system which does not contain the MZMs.
What is more important, the presented method correctly reproduces the spectrum of the system.
Because the algorithm is based on the Krylov method, the time of eigenstates computation is proportional to the size of the system.
Our method is not limited by any specific form of a  noninteracting Hamiltonian and can be applied to a system with arbitrary shape or geometry.

The main advantage of our method is that the topological in-gap states can be identified via looking for extreme eigenvalues and the corresponding eigenvectors of a semi-positive definite, symmetric matrix. 
Due to the latter property, one may apply the Lanczos algorithm and study tight-binding models containing up to $10^6$ lattice sites (for a single band model). 
This opens a new way to study the edge states in the topologically nontrivial systems of relatively large sizes.
We have compared numerical results obtained via the presented method and within the Bogoliubov--de~Gennes technique and we have found that both methods are in excellent agreement (cf. Sec.~\ref{sec.basic_example}).
Additionally, we have applied our method to a few commonly studied models and compared numerical results with the results from the literature---e.g. for the 2D Kitaev model, 2D Rashba stripes, or 3D Rashba nanowire (cf. Sec.~\ref{sec.adv_example}).

Finally, we have applyied the presented method to study the realization of the topological phase in the magnetic nanoisland deposited at the superconducting surface (cf. Sec.~\ref{sec.island}).
This study was motivated by recently realized experiments  ~\cite{menard.guissart.17,palaciomorales.mascot.19}, in which the realization of the non-trivial topological phase is observed in form of the edge states. 
We have shown that the eigenstates of the system, depends on the length of the edge of the nanoisland and in the thermodynamic limit
tend to zero.

Presented comparative analysis shows that the discussed method can be successfully applied for different types of systems (independently of its dimension, geometry, shape, etc.).
Finally, we note that our method has an additional important property. 
Namely, it targets not only the Majorana zero modes but general problem of the topological in-gap states. 
From this, our method can be applied to study of various type of physical phenomena, like Majorana zero modes, localized in-gap states, or edge states in high order topological insulators.

\begin{acknowledgments}
We thank Pascal Simon for inspiriting discussions.
This work was supported by the National Science Centre (NCN, Poland) under grants:
2016/23/B/ST3/00647 (A.W. and M.M.), %MM
2017/24/C/ST3/00276 (A.P.), %AP
and
2019/33/N/ST3/03137 (M.K.). %MK
A.P. appreciates funding in the frame of scholarships of the Minister of Science and Higher Education (Poland) for outstanding young scientists (2019 edition, no. 818/STYP/14/2019).
\end{acknowledgments}

%%%%%%%%%%%%%%%%%%%%%%%%%%%%%%%%%%%%%%%%%%%%%
%%%%%%%%%%%%%%%%%%%%%%%%%%%%%%%%%%%%%%%%%%%%%
%%%%%%%%%%%%%%%%%%%%%%%%%%%%%%%%%%%%%%%%%%%%%

\appendix

\section{Details of the method}
\label{app.details}
In this Appendix, we present the technical details of derivations in
Section~\ref{sec.method}.
The properties of the general Majorana Hamiltonian~(\ref{eq:general_Majorana_hamiltonian}) are discussed in the Appendix~\ref{app.details.M_matrix}.
The Appendix~\ref{app.details.derviation} shows how to derive our central result~(\ref{eq.Majorana_matrix_condition}) starting from Eq.~(\ref{eq.Majorana_condition}).

\subsection{Properties of Hamiltonian in the base of Majorana operator}
\label{app.details.M_matrix}

The matrix ${M}_{ij}$ in Hamiltonian (\ref{eq:general_Majorana_hamiltonian}) is not uniquely defined, because of the commutation relations. 
For any arbitrary pair of indexes $i,j$ the following transformation can be done
\begin{eqnarray}
    {M}_{ij}\gamma_i \gamma_j + {M}_{ji}\gamma_j \gamma_i = 
    \left( {M}_{ij} - {M}_{ji} \right) \gamma_i \gamma_j.
\end{eqnarray}
From above transformation it is easy to see that the diagonal elements of that matrix are irrelevant and may be eliminated.
Thus, without loss of generality, this matrix can be uniquely chosen in the upper triangular form
\begin{eqnarray}
M_{ij} = \begin{cases} {M}_{ij} & i<j\\
            0, & i \geq j  \end{cases}
\end{eqnarray}

Since the Hamiltonian as well as the Majorana operators are Hermitian, then the matrix $M$ is real,
\begin{eqnarray}
\text iM_{ij} \gamma_i \gamma_j =  \left(\text iM_{ij} \gamma_i \gamma_j \right)^\dagger =  \text iM^*_{ij} \gamma_i \gamma_j.
\end{eqnarray}

\subsection{Details of the condition for MZMs}
\label{app.details.derviation}

The Eq.~(\ref{eq.Majorana_condition}) can be expanded in the following form,
\begin{eqnarray}
\label{eq.Majorana_condition_extended}
\sum_m \left(h_{nm}-h_{mn} \right)^2 = \sum_m \left( h^2_{nm}+h^2_{mn}-2h_{nm}h_{mn}  \right). 
\end{eqnarray}
Three terms: $\sum_m h^2_{nm}$, $\sum_m h^2_{mn}$, $\sum_m 2h_{nm}h_{mn}$ are calculated in the next three Equations: (\ref{eq.h_nm}), (\ref{eq.h_mn}), and (\ref{eq.h_nm.h_mn}), respectively:
\begin{eqnarray}
\label{eq.h_nm}
\sum_m h^2_{nm} 
&=& \sum_{i,i',j} O^T_{ni} M_{ij} M^T_{j,i'}  O_{i',n} 
\end{eqnarray}
\begin{eqnarray}
\label{eq.h_mn}
\sum_m h^2_{mn} 
&=& \sum_{i,j,j'} O^T_{nj} M^T_{ji}  M_{i,j'}  O_{j',n} 
\end{eqnarray}
\begin{eqnarray}
\label{eq.h_nm.h_mn}
\nonumber 2 \sum_m h_{nm} h_{mn} &=& \sum_{i,i',j} O^T_{ni} M_{ij} M_{j,j'} O_{j'n} + \sum_{i,i',j} O^T_{ni} M^T_{ij} M^T_{j,j'} O_{j'n} \\
\end{eqnarray}
By the substitution %Eq.~(\ref{eq.h_nm}), (\ref{eq.h_mn}), (\ref{eq.h_nm.h_mn}) 
Eq.~(\ref{eq.h_nm})--(\ref{eq.h_nm.h_mn}) to the Eq. (\ref{eq.Majorana_condition_extended}), the following result is obtained, 
\begin{eqnarray}
&& \sum_m \left(h_{nm}-h_{mn} \right)^2 \\
\nonumber &=& \sum_{i,j,k} O^T_{ni}\left( M_{ij} M^T_{j,k} + M^T_{ij} M_{j,k} - M^T_{ij} M^T_{j,k} - M_{ij} M_{j,k}  \right) O_{k,n}. 
\end{eqnarray}
Thus, the final condition for MZMs can be written in the next compact form,
\begin{eqnarray}
\sum_{i,j} O^T_{ni} \left[(M-M^T)^2\right]_{ij} O_{j,n} = 0.
\end{eqnarray}

\section{Bogoliubov--de~Gennes equations technique}
\label{sec.bdg_details}

In the general case, the Hamiltonian $\mathcal{H}$ describing the finite size system in real space, can be diagonalized by unitary transformation~\cite{balatsky.vekhter.06}
\begin{eqnarray}
a_{i\sigma} &=& \sum_{n} \left( u_{in\sigma} \eta_{n} 
- \sigma v_{in\bar{\sigma}}^{\ast} \eta_{n}^{\dagger} \right) , 
\label{eq.bvtransform}
\end{eqnarray}
where $\eta_{n}$ nad $\eta_{n}^{\dagger}$ are the new fermionic annihilation and creation operators.
This transformation leads to {\it the Bogoliubov--de~Gennes equations} in the form $E_{n} \Psi_{in} = \sum_{j} \mathbb{H}_{ij} \Psi_{jn}$, where
\begin{eqnarray}
\mathbb{H}_{ij} &=& \left(
\begin{array}{cccc}
H_{ij\uparrow\uparrow} & H_{ij\uparrow\downarrow} & D_{ij}^{s} & D_{ij}^{t} \\ 
H_{ij\downarrow\uparrow} & H_{ij\downarrow\downarrow} & D_{ij}^{t} & D_{ij}^{s} \\ 
( D_{ij}^{s} )^{\ast} & ( D_{ij}^{t} )^{\ast} & -H_{ij\downarrow\downarrow}^{\ast} & -H_{ij\downarrow\uparrow}^{\ast} \\
( D_{ij}^{t} )^{\ast} & ( D_{ij}^{s} )^{\ast} & -H_{ij\uparrow\downarrow}^{\ast} & -H_{ij\uparrow\uparrow}^{\ast}
\end{array} \right) \label{eq.bdg}
\end{eqnarray}
is the matrix form of the Hamiltonian $\mathcal{H}$, while eigenvectors are given by
\begin{eqnarray}
\Psi_{in} &=& \left( u_{in\uparrow} , u_{in\downarrow} , v_{in\downarrow} , v_{in\uparrow} \right)^{T} .
\end{eqnarray}
Here, the block matrices denote: $H_{ij\sigma\sigma}$ free-electron term (in the form of spin-conserve hopping term), $H_{ij\sigma\bar{\sigma}}$ spin--orbit coupling (in the form of spin-flip hopping terms), while $D_{ij}^{s}$ and $D_{ij}^{t}$ spin-singlet and spin-triplet superconducting term.
For instant, in the case of described 1D lattice with {\it s-wave} on-site superconductivity (Sec.~\ref{sec.basic_example}), we have $H_{ij\sigma\sigma} 
= t \delta_{\langle i,j \rangle} + ( \mu + \sigma h ) \delta_{ij}$,
$H_{ij\sigma\sigma'} = -\text i \alpha \sigma^{y}_{\sigma\sigma'} \left( \delta_{i+1,j} - \delta_{i-1,j} \right)$, and $D_{ij}^{s} = \Delta \delta_{ij}$ while $D_{ij}^{t} = 0$.
From solution of the BdG equation, one can determine measurable physical quantities, e.g. the local density of states (LDOS)~\cite{matsui.sato.03}
\begin{eqnarray}
\rho_{i} (\omega) &=& \sum_{n\sigma} \left[ | u_{in\sigma} |^{2} \delta(\omega + \mathcal{E}_{n}) + | v_{in\sigma} |^{2} \delta(\omega - \mathcal{E}_{n}) \right] ,
\label{eq.bdg_ldos}
\end{eqnarray}
where $\delta(\omega)$ denotes the Dirac delta function. 
In numerical calculations we replace the delta function by the Lorentzian $\delta(\omega) = \zeta/[\pi (\omega^2+\zeta^{2})]$, with a small broadening $\zeta/t = 0.001$.

\bibliography{bib.bib}

\end{document}